\documentclass[journal,a4paper]{IEEEtran}

\usepackage{cite}
\usepackage{algorithm}
\usepackage{algorithmicx}
\usepackage{algpseudocode}
\usepackage{balance}
\usepackage{graphicx}
\usepackage{tabularx}
\usepackage{epstopdf}
\usepackage{url}
\usepackage{amssymb}
\usepackage{amsthm}
\usepackage{tikz}
\usepackage{textcomp}
\usepackage{marvosym}
\usepackage{subfig}
\usepackage{newtxmath}
\usepackage{enumerate}
\usepackage{mathtools,amsthm}

\newtheorem{theorem}{Theorem}
\newtheorem{lemma}{Lemma}

\newtheorem{assumption}{Assumption}

\newtheorem{remark}{Remark}

\DeclareMathOperator{\col}{col}
\DeclareMathOperator{\sgn}{sgn}
\DeclareMathOperator{\rank}{rank}
\DeclareMathOperator{\tr}{tr}
\DeclareMathOperator{\sig}{sig}

\begin{document}

\title{Distributed Continuous-Time Optimization with Uncertain Time-Varying Quadratic Cost Functions}

\author{Liangze Jiang, Zheng-Guang Wu, and Lei Wang
\thanks{The authors are with College of Control Science and Engineering, Zhejiang University, Hangzhou 310027, China (e-mail: zetojiang; nashwzhg; lei.wangzju @zju.edu.cn). (\em Corresponding author: Zheng-Guang Wu.)}}

\markboth{}%
{Shell \MakeLowercase{\textit{et al.}}: Bare Demo of IEEEtran.cls for IEEE Journals}

\maketitle

\begin{abstract}
This paper studies distributed continuous-time optimization for time-varying quadratic cost functions with uncertain parameters. We first propose a centralized adaptive optimization algorithm using partial information of the cost function. It can be seen that even if there are uncertain parameters in the cost function, exact optimization can still be achieved. To solve this problem in a distributed manner when different local cost functions have identical Hessians, we propose a novel distributed algorithm that cascades the fixed-time average estimator and the distributed optimizer. We remove the requirement for the upper bounds of certain complex functions by integrating state-based gains in the proposed design. We further extend this result to address the distributed optimization where the time-varying cost functions have nonidentical Hessians. We prove the convergence of all the proposed algorithms in the global sense. Numerical examples verify the proposed algorithms.
\end{abstract}


\section{Introduction}
\label{Introduction}
Distributed control of large-scale multi-agent systems has been an enduring topic\cite{Olfati,Guilong24,GUO23}. Recently, based on optimization and game theories, exploring new cooperative control methods within the framework of decision-making problems, such as distributed online optimization\cite{Simonetto20,LI2023100904} and distributed learning-based control\cite{Bing24,Yongliang24}, has received increasing attention. We are interested in the distributed optimization, where a group of agents searches for the optimal solution to a global cost function through the information exchange between agents and their neighbors. Early relevant research can be traced back to \cite{Nedic09,wang2010control,gharesifard2013distributed}, where the combination of consensus and subgradient methods provides a basic approach for solving distributed optimization problems. Subsequent advancements have expanded the algorithms from fixed to flexible information exchange topology\cite{yang2016distributed,Chen21}. The dynamics of controlled agents has been generalized from single- or double-integrators to high-order dynamics and Euler-Lagrange systems\cite{Tang19,Wang21,QIN2022110113}. Moreover, event-triggered communication\cite{kia2015}, collision avoidance, and formation behaviors during the process of finding the optimal solution \cite{An21collision,Jiang22} have also been further considered. 

The studies mentioned above require the parameters of the cost functions are time-invariant. However, the cost functions modeling in many engineering systems may exhibit time-varying natures\cite{su2009traffic,Simonetto20}. In these cases, directly using algorithms suitable for time-invariant optimization problem may introduce tracking error to the time-varying optimal solution. By further introducing the prediction dynamics in the optimization algorithm, the tracking error can be eliminated exactly \cite{su2009traffic,Simonetto16,Fazlyab18}. This basic idea is also expanded to address distributed time-varying optimization problem\cite{Ren17}. Then, the time-varying optimization problem for a class of nonlinear multi-agent systems is considered in \cite{Huang20} by employing an auxiliary estimator and a tracking controller. With the help of output regulation theory, the periodically time-varying unconstrained and constrained optimization problems are solved in \cite{Ding22}. Based on the finite-time average estimator, the resource allocation problem with time-varying quadratic cost functions is studied in \cite{wang2022distributed}.

However, in some practical scenarios, the parameters in the cost functions may not always be available. For example, in signal processing problems, data information is provided by measurements from sensor networks. Therefore, the time-varying nature of parameters comes from the dynamic nature of the monitoring process or unknown sensor failures\cite{Bastianello24}. In motor control problems, to reduce torque ripple of permanent magnet synchronous motors, the harmonic current injection function should be minimized, but this function contains at least four parameters, which are difficult to obtain accurately through conventional methods \cite{yan2019torque}. Moreover, in source seeking problems \cite{fabbiano2014source,kim2014cooperative,brinon2019multirobot}, the establishment of the cost function depends on the moving signal source. Therefore, if the power and trajectory parameters of the mobile signal source cannot be obtained in advance, this in turn leads to uncertainty in the parameters of the cost function (see details in Section \ref{problem}). In the case where there are unknown (or inexact) parameters in the cost functions caused by unknown dynamic environments, most existing algorithms can not be applied directly since the known parameters of the cost functions are needed for exact optimization. Few results have been obtained for this problem, such as the centralized discrete-time design in \cite{Bastianello24}.

Motivated by this, we  study the distributed continuous-time optimization for time-varying quadratic cost functions, which contains the uncertain parameters. We first focus on the design of a centralized adaptive optimization algorithm by employing {\em partial} information of the cost function, and then study its distributed implementation. Based on a structure that cascades the fixed-time average estimator and the distributed optimizer, we propose two distributed algorithms for cases where the Hessians of all local cost functions are identical and nonidentical, respectively. The contributions of the paper include:

\begin{itemize}
    \item The proposed algorithms achieve the exact optimization with the existence of uncertain parameters in the cost functions, which relax the assumption that the cost function is fully known\cite{su2009traffic,Simonetto16,Fazlyab18,Ren17,Sun17,Huang20,wang2022distributed,Ding22}.
    
    \item By introducing state-based gains in the estimator and optimizer, the requirement for the upper bounds of complex time-varying functions is removed, which is different from \cite{Ren17,Sun17,Chu22,wang2022distributed}. Moreover, the convergence of the proposed algorithms is established in the global sense.
    \item Since the discontinuous function only exists in the estimator's dynamics and the time-varying continuous approximation method is applied to discontinuous function in the optimizer, the chattering phenomenon of the closed-loop output is greatly reduced compared with \cite{Ren17,wang2022distributed} and \cite{Sun23}. 
\end{itemize}
{\bf Notations.} The set of real and nonnegative numbers are represented by $\mathbb{R}$ and $\mathbb{R}_+$, respectively. Denote $1_N$ the vector $\col(1,\cdots,1)\in\mathbb{R}^N$ and $I_m$ the $m$-dimensional identity matrix. For a vector or a matrix $Y$, we use $||Y||_p$ to represent the $l_p$-norm of $Y$. For a matrix $A\in\mathbb{R}^m$, we use $\tr(A)$ to represent the trace of $A$. For a real number $y\in\mathbb{R}$, the signum function is denoted by $\sgn(y)$. We define $\sig(y)^\alpha=|y|^\alpha\sgn(y)$ with $y\in\mathbb{R}$ and $\alpha>0$. Define a nonlinear function $S(y,\epsilon_1)=y/(|y|+(1/\epsilon_1) \exp(-ct))$ with $y\in\mathbb{R}$ and $\epsilon_1,c>0$. For a vector $x=\col(x_1,\cdots,x_n)\in\mathbb{R}^n$, define $\sgn(x)=\col(\sgn(x_1),\cdots,\sgn(x_n))$, $S(x,\epsilon_1)=\col(S(x_1,\epsilon_1),\cdots,S(x_n,\epsilon_1))$ and $\sig(x)^\alpha=\col(\sig(x_1)^\alpha,\cdots,$ $\sig(x_N)^\alpha)$. For a differentiable function $f(x,t):\mathbb{R}^n\times\mathbb{R}_+\rightarrow \mathbb{R}$, we use $\nabla f(x,t)$ to represent the gradient of $f(x,t)$ with respect to $x$, $\nabla_{t} f(x,t)$ to represent the partial derivative of $\nabla f(x,t)$ with respect to $t$, and $\nabla_{x} f(x,t)$ to represent the Hessian of $\nabla f(x,t)$ with respect to $x$.

\begin{lemma}\label{finite}
\cite{Polyakov12} Consider the system $\dot{x}=g(x,t)$, $x(0)=x_0$, where $x\in\mathbb{R}^m$ and $g:\mathbb{R}^n\times\mathbb{R}_+\rightarrow\mathbb{R}^m$ is a nonlinear function. Suppose there exists a continuous function $V:\mathbb{R}^m\rightarrow\mathbb{R}$ and some positive constants $a,b>0$ and $0<\mu_1<1<\mu_2$ such that $\dot{V}(x)\leq-a(V(x))^{\mu_1}-b(V(x))^{\mu_2}$, then the origin is globally fixed-time stable for $\dot{x}=g(x,t)$. This means that all solutions satisfy $\lim_{t\rightarrow \infty} x(t)=0$ and $x(t)=0$, $\forall t\geq t_0$, where $t_0\leq1/a(1-\mu_1)+1/b(\mu_2-1)$.
\end{lemma}
\begin{lemma}\label{fangda}
\cite{ni2019new} Let $x_1$, $x_2$, $\cdots$, $x_N\geq0$, and $p>1$. Then,
$\sum_{i=1}^{N}x_i^p\geq N^{1-p}\left(\sum_{i=1}^{N}x_i\right)^p$.
\end{lemma}

\section{Preliminaries and Problem Statement}
\label{Preliminaries and problem statement}

\subsection{Preliminaries}

A differentiable function $f(x,t):\mathbb{R}^n\times\mathbb{R}_+\rightarrow \mathbb{R}$ is uniformly convex in $x$ if, for all $a$, $b$ in $\mathbb{R}^n$ and for all $t\geq 0$, the inequality $f(a,t)-f(b,t)\geq \nabla f(b,t)^{\top}(a-b)$ holds. The function $f(x,t)$ is uniformly $m$-strongly convex in $x$ if, for all $a$, $b$ in $\mathbb{R}^n$ and for all $t\geq 0$, there exists a positive constant $m$ such that $(a-b)^{\top}(\nabla f(a,t)-\nabla f(b,t))\geq m||a-b||^2$ holds.

The information exchange topology is modeled as an undirected graph $\mathcal{G}=(\mathcal{N},\mathcal{E})$, where $\mathcal{N}=\left\{1,\cdots,N\right\}$ represents the set of nodes corresponding to agents, and $\mathcal{E}\subseteq \mathcal{N}\times \mathcal{N}$ is the set of edges. The graph's weighted adjacency matrix is $\mathcal{A}=[a_{ij}]_{N\times N}$. An edge $(i,j)\in \mathcal{E}$ indicates that nodes $i,j$ can exchange information with each other. If $(j,i)\in \mathcal{E}$, then $a_{ij}>0$, and $a_{ij}=0$ otherwise. The graph's Laplacian, $L=[l_{ij}]_{N\times N}$, is defined as $l_{ii}=\sum_{j\neq i}a_{ij}$ and $l_{ij}=-a_{ij}$ for $j\neq i$. A graph $\mathcal{G}$ is connected if every node is reachable from any other node. By assigning an orientation for the edges in graph $\mathcal{G}$, denote the incidence matrix associated with $\mathcal{G} $ as $D=[d_{ik}]_{N\times |\mathcal{E}|}$, where $d_{ik}=1$ if edge $e_k$ enters node $i$, $d_{ik}=-1$ if it leaves, and $d_{ik}=0$ otherwise. 

\begin{assumption}\label{assumption.graph1}
The graph $\mathcal{G}$ is undirected and connected.
\end{assumption}

According to this assumption, we know that the graph's Laplacian matrix $L$ features a zero eigenvalue corresponding to the $1_N$ eigenvector, and the other eigenvalues are positive. We order these eigenvalues as $\lambda_1=0<\lambda_2\leq\cdots\leq\lambda_N$.


\subsection{Problem Definition}\label{problem}

Consider a group of $N$ agents with an information exchange topology $\mathcal{G}=(\mathcal{N},\mathcal{E})$. Each agent $i\in\mathcal{N}$ is modeled as the single-integrator dynamics
\begin{align}\label{eq.multiagent}
    \dot{x}_i(t)=u_i(t),
\end{align}
where $x_i\in\mathbb{R}^m$ and $u_i\in\mathbb{R}^m$ are agent $i$'s state (e.g., generalized coordinate) and control input. Suppose each agent is assigned a local cost function
\begin{align}\label{eq.fi}
    f_i(x_i,t)=\frac{1}{2}x_i^{\top}H_i(t)x_i+R_i^{\top}(t)x_i+d_i(t).
\end{align}
It satisfies $\nabla_{x_i} f_i(x_i,t)=H_i(t)=\Omega_ih_i(t)$ and $\nabla_{t} f_i(x_i,t)=\dot{H}_i(t)x_i+\dot{R}_i(t)=A_ig_i(x_i,t)$, where $\Omega_i\in \mathbb{R}^{m\times m}$ and $A_i\in \mathbb{R}^{m\times p_i}$ are {\em uncertain} constant matrices, and $h_i(t)\in \mathbb{R}^{m\times m}$ and ${g}_i(x_i,t)\in \mathbb{R}^{p_i}$ are known matrix-valued and vector-valued functions, respectively. The objective of this paper is to design a proper $u_i(t)$ such that the states of all agents converge to the solution of the optimization problem
\begin{align}
&\min_{x\in\mathbb{R}^{mN}}f(x,t),~~~f(x,t)=\sum_ {i\in \mathcal{N}}{f_i(x_i,t)},\nonumber\\
&~~s.t.~~x_i=x_j~~~~~ i,j\in\mathcal{N},\label{optimizationproblem}
\end{align}
where $x=\col(x_1,\dots,x_N)$, and $f(x,t)$ is the global cost function. Denote $x^*(t):=\arg\min_{x\in\mathbb{R}^m}$ $f(x,t)$. It is worth noting that in our problem setting,
\begin{enumerate}[(1)]
    \item Both functions $\nabla_{x_i} f_i(x_i,t)$ and $\nabla_{t} f_i(x_i,t)$ can be parameterized as a matrix multiplied by a known matrix-valued or vector-valued function. Many cost functions satisfy this property, such as the cost function in the following {\bf Example 1}.
    \item Agent $i$ is allowed to use the {\em real-time measurable gradient value} $\nabla f_i(x_i,t)$ at its position $x_i$ at time instant $t$\cite{Tang19,QIN2022110113,Bastianello24}, rather than the gradient function $\nabla f_i(x_i,t)$, which provides the value of $\nabla f_i(x_i,t)$ for any $x_i$ and $t$. This is because the existence of uncertain parameters in cost functions renders the analytical $\nabla f_i(x_i,t)$ unavailable.
\end{enumerate}

{\bf Example 1 (A Specific Example)}: Consider an example where an autonomous robot searches for a moving signal source \cite{fabbiano2014source}. The signal strength received by the robot is modeled as 
\begin{align}\label{eq.modelstrength}
   P(y,t)=\frac{a(t)}{||y-r(t)||^2},~~~~~y\neq r(t),
\end{align}
where $a(t)>0$ and $r(t)\in\mathbb{R}^{m}$ denote the power and the position of the source at time $t$, respectively, and $y\in\mathbb{R}^m$ is the position of the robot. This problem of searching a signal source can be  mathematically described as an optimization problem of finding the minimum of a cost function, i.e., $\min_{y}P^{-1}(y,t)=a^{-1}(t)||y-r(t)||^2$, whose optimal solution is exactly $r(t)$. Assume that each agent can obtain the gradient value corresponding to its local cost function through the {\em real-time measurement} methods\cite{fabbiano2014source,brinon2019multirobot}. Clearly, if the function $P^{-1}(y,t)$ is fully known, the existing time-varying optimization algorithms\cite{su2009traffic,Ren17} can be applied directly to solve it. However, once there are parameters in $P^{-1}(y,t)$ that cannot be obtained in advance\cite{kim2014cooperative}, such as the function $P^{-1}(y,t)=k_1||y-k_2t||^2$ with unknown $k_1>0$ and $k_2\in\mathbb{R}$, the search for the optimal solution becomes a challenging problem. Motivated by this, the following sections explore novel algorithms that achieve exact optimization when there exist uncertain parameters in the cost functions. 

We make the following assumptions on the cost functions.

\begin{assumption}\label{assumption.localcostfunction}
For any $i\in\mathcal{N}$, $f_i(x_i,t)$ is uniformly $\bar{H}_1$-strongly convex with respect to $x_i$, $\forall t\geq 0$.
\end{assumption}
Assumption \ref{assumption.localcostfunction} means that $H_i(t)$ is invertible and thus, $\sum_{i\in\mathcal{N}}H_i(t)$. Then, there exists a unique solution $x^*(t)$ for the optimization problem (\ref{optimizationproblem}).
\begin{assumption}\label{assumption.boundedness}
For any $i\in\mathcal{N}$, $||{R}_i(t)||_\infty$ is bounded. Moreover, there exist positive constants $\bar{H}_2$ and $\bar{R}$ such that $||H_i(t)||_\infty\leq\bar{H}_2$, $||\dot{H}_i(t)||_\infty\leq\bar{H}_2$ and $||\dot{R}_i(t)||_\infty\leq\bar{R}$, $\forall t\geq 0$.
\end{assumption}
It can be concluded from Assumption \ref{assumption.boundedness} that the local optimal trajectory $x^*_i(t)$, and its time derivative $\dot{x}^*_i(t)$ are bounded by using the optimal condition $\nabla f_i(x^*_i(t),t)=0$, $\forall t\geq 0$. Similar assumptions on strong convexity and parameter boundedness have also been adopted by \cite{Ren17,Sun17,Chu22,GUO24}. 
\begin{remark}
Considering that the parameter functions may be nonsmooth in dynamic environments \cite{Santilli24}, Assumption \ref{assumption.boundedness} can be further extended to address globally Lipschitz parameter functions. Therefore, the concepts of generalized gradient \cite{Clarke1987Optimization} and nonsmooth analysis \cite{Cortes08} should be used to analyze the convergence of the solution. Since the Lyapunov function employed in the analysis is continuously differentiable, the subsequent proofs still hold.
\end{remark}

\section{Design From a Centralized Perspective}
\label{cen}
We first consider the centralized time-varying optimization
\begin{align}\label{centralizedoptimization}
    \min_{y\in\mathbb{R}^m}f_c(y,t).
\end{align}
The function $f_c(y,t):\mathbb{R}^m\times \mathbb{R}_+\rightarrow \mathbb{R}$ conforms to the form of \eqref{eq.fi} with time-varying $H_c(t)$ and $R_c(t)$, and satisfies
\begin{align}\label{eq.rewritehp}
    \nabla_yf_c(y,t)=\Omega_ch_c(t),~~~\nabla_t f_c(y,t)=A_cg_c(y,t). 
\end{align}
Denote $y^*(t):=\mathop{\arg\min}_{y\in\mathbb{R}^m}f_c\left(y,t\right)$. The dynamics of the agent is modeled as
\begin{align}\label{singleagent}
    \dot{y}(t)=u(t).
\end{align}
To avoid symbol redundancy, we use $H_c$ to represent $H_c(t)$, $h_c$ to represent ${h}_c(t)$, and $g_c$ to represent $g_c(y,t)$ in the subsequent analysis. Notations $H_i(t)$, $h_i(t)$ and $g_i(x_i,t)$ in the following sections are simplified similarly. 

Before giving the design of $u(t)$ to solve (\ref{centralizedoptimization}), we show the reversibility of $\Omega_c$ and $h_c$, which is important in the following design. For any matrices $X$ and $Y$ with compatible dimensions, we know from \cite{matrix} that $\rank(XY)\leq\min\{\rank{(X)},\rank{(Y)}\}$. By Assumption \ref{assumption.localcostfunction}, the Hessian $H_c$ is invertible, which results in $\min\{\rank{(\Omega_c)},\rank{(h_c)}\}\geq m$. With the fact that $\rank{(\Omega_c)}\leq m$ and $\rank{(h_c)}\leq m$, we know $\rank{(\Omega_c)}=\rank{(h_c)}=m$ which further implies the reversibility of $\Omega_c$ and $h_c$, $\forall t\geq0$. Based on this observation, the centralized optimization algorithm is given by
\begin{align}
    u=&-k_c\nabla f_c(y,t)-h_c^{-1}\hat{\eta}_1{g}_c,\label{eq.f.controla}\\
    \dot{\hat{\eta}}_1=&\gamma_1(h_c^{-1})^{\top}\nabla f_c(y,t){g}_c^{\top},\label{eq.f.controlb}
\end{align}
where $k_c$ is a positive constant to be determined later, $\hat{\eta}_1\in \mathbb{R}^{m\times p}$ is an adaptive gain matrix, and $\gamma_1\in\mathbb{R}^{m\times m}$ is a positive definite constant matrix which can be adjusted to improve transient performance \cite{KKKbook} of closed-loop system.
\begin{remark}
    If parameters $\Omega_c$ and $A_c$ are known in advance, then both $\nabla_y f_c(y,t)$ (i.e. $H_c$) and $\nabla_t f_c(y,t)$ are available, and thus, optimization problem (\ref{centralizedoptimization}) can be solved readily by using the algorithm given by \cite{su2009traffic} with
    \begin{align}\label{eq.primal}
        u=-\nabla f_c(y,t)-H_c^{-1}\nabla_t f_c(y,t).
        \end{align}
    Compared with this design, the proposed design \eqref{eq.f.controla}-\eqref{eq.f.controlb} replaces $-H_c^{-1}\nabla_t f_c(y,t)$ with $-h_c^{-1}\hat{\eta}_1{g}_c$ as an estimated `feedforward' term.
\end{remark}

\begin{theorem}\label{centralized.oneorder}
Suppose that the function $f_c(y,t)$ satisfies Assumptions \ref{assumption.localcostfunction}-\ref{assumption.boundedness}. For the system \eqref{singleagent}-\eqref{eq.f.controlb}, by selecting $k_c>\sqrt{m}\bar{H}_2/{(2\bar{H}_1^2)}$, the optimization problem (\ref{centralizedoptimization}) is solved, i.e., $\lim_{t\rightarrow \infty} y(t)-y^*(t)=0$.
\end{theorem}

\begin{proof}
Since that parameters $\Omega_c$ and $A_c$ are not available, we use an adaptive gain matrix $\hat{\eta}_1$ to estimate the coupled gain matrix $\Omega_c^{-1}A_c$. By defining the estimation error
\begin{align}\label{eq.tildeeta}
    \tilde{\eta}_1={\Omega_c}^{-1}A_c-\hat{\eta}_1,
\end{align}
consider the following Lyapunov function candidate 
\begin{align}
    V(y,\tilde{\eta}_1,t)=\frac{1}{2}\nabla f_c(y,t)^{\top}H_c^{-1}\nabla f_c(y,t)+\frac{1}{2}\tr(\tilde{\eta}_1^{\top}\gamma_1^{-1}\tilde{\eta}_1).\nonumber
\end{align}
Its derivative is given by
\begin{align}
    \dot{V}=&\nabla f_c(y,t)^{\top}H_c^{-1}\frac{d}{dt}{\nabla} f_c(y,t)\nonumber\\
    &+\frac{1}{2}\nabla f_c(y,t)^{\top}\left(\frac{d}{dt}H_c^{-1}\right){\nabla} f_c(y,t)+\tr(\tilde{\eta}_1^{\top}\gamma_1^{-1}\dot{\tilde{\eta}}_1).
\end{align}
Note that 
\begin{align}\label{eq.Delta1}
&\frac{1}{2}\bigg|\bigg|\nabla f_c(y,t)^{\top}\left(\frac{d}{dt}H_c^{-1}\right){\nabla} f_c(y,t)\bigg|\bigg|\nonumber\\
=&{\frac{1}{2}\bigg|\bigg|\nabla f_c(y,t)^{\top}\left(H_c^{-1}\dot{H}_cH_c^{-1}\right){\nabla} f_c(y,t)\bigg|\bigg|}\nonumber\\
{\leq}&{\frac{\sqrt{m}\bar{H}_2}{2\bar{H}_1^2}||{\nabla} f_c(y,t)||^2=:\Delta_1}.
\end{align}
We then derive that
\begin{align}
\dot{V}\leq&\nabla f_c(y,t)^{\top}\dot{y}+\nabla f_c(y,t)^{\top}H_c^{-1}\nabla_t f_c(y,t)\nonumber\\
&+\tr(\tilde{\eta}_1^{\top}\gamma_1^{-1}\dot{\tilde{\eta}}_1)+{\Delta_1}\nonumber\\
=&-{k_c}||\nabla f_c(y,t)||^2+\nabla f_c(y,t)^{\top}(H_c^{-1}A_c-h_c^{-1}\hat{\eta}_1){g}_c\nonumber\\
&+\tr(\tilde{\eta}_1^{\top}\gamma_1^{-1}\dot{\tilde{\eta}}_1)+{\Delta_1}\nonumber\\
=&-{k_c}||\nabla f_c(y,t)||^2+\nabla f_c(y,t)^{\top}{h_c}^{-1}\tilde{\eta}_1{g}_c\nonumber\\
&-\tr(\tilde{\eta}_1^{\top}\gamma^{-1}\dot{\hat{\eta}}_1)+{\Delta_1},
\end{align}
where the second equation follows from \eqref{eq.f.controla}, and the third equation follows from \eqref{eq.rewritehp} and \eqref{eq.tildeeta}. According to the properties of trace of a matrix \cite{matrix}, we have 
\begin{align}\label{eq.dotv}
    \dot{V}\leq&-{k_c}||\nabla f_c(y,t)||^2+\tr\left(h_c^{-1}\tilde{\eta}_1g_c\nabla f_c(y,t)^{\top}\right)\nonumber\\
    &-\tr(\tilde{\eta}_1^{\top}\gamma^{-1}\dot{\hat{\eta}}_1)+{\Delta_1}\nonumber\\
    =&-{k_c}||\nabla f_c(y,t)||^2+\tr\left(\tilde{\eta}_1^{\top}({h_c}^{-1})^{\top}\nabla f_c(y,t){g}_c^{\top}\right)\nonumber\\
    &-\tr(\tilde{\eta}_1^{\top}\gamma_1^{-1}\dot{\hat{\eta}}_1)+{\Delta_1}\nonumber\\
    =&-{\left(k_c-\frac{\sqrt{m}\bar{H}_2}{2\bar{H}_1^2}\right)}||\nabla f_c(y,t)||^2,
\end{align}
where the first equation follows from the fact that $\tr(ba^{\top})=a^{\top}b$ for any compatible vectors $a$ and $b$, the second equation follows from $\tr(X^{\top}Y)=\tr(XY^{\top})$ for any compatible matrices $X$ and $Y$, and the last equation follows from \eqref{eq.f.controlb}. Thus, if $k_c>\sqrt{m}\bar{H}_2/{(2\bar{H}_1^2)}$ holds, we have $V(y,\tilde{\eta}_1,t)\leq V(y(0),\tilde{\eta}_1(0),0)$, $\forall t\geq 0$, which implies the boundedness of $\nabla f_c(y,t)$ and $\tilde{\eta}_1$. From \eqref{eq.tildeeta}, $\hat{\eta}_1$ is bounded. By integrating both sides of \eqref{eq.dotv}, we have $\nabla f_c(y,t)\in\mathcal{L}_2$. We next show the boundedness of $(d/dt)\nabla f_c(y,t)$. According to \eqref{eq.fi}, we have $\nabla f_c(y,t)=H_cy+R_c$. Combining the boundedness of $\nabla f_c(y,t)$ and the boundedness of $H_c$ and $R_c$ given by Assumption \ref{assumption.boundedness}, we know $y$ is bounded. With the fact that $\nabla_t f_c(y,t)=\dot{H}_cy+\dot{R}_c=A_cg_c$, the bounded $y$ guarantees the boundedness of $\nabla_t f_c(y,t)$, and thus $g_c$ is bounded. By recalling \eqref{eq.f.controla}, $(d/dt)\nabla f_c(y,t)$ is bounded. Then, by applying Barbalat'Lemma \cite{Nonlinear02khalil}, we conclude that $\lim_{t\rightarrow\infty}\nabla f_c(y,t)=0$. With the strong convexity of $f_c(y,t)$, there exists a positive constant $m_c$ such that for any $x,y\in \mathbb{R}^m$,
\begin{align}\label{eq.inequalitystrongconvex}
    m_c||y-x||\leq||\nabla f_c(y,t)-\nabla f_c(x,t)||. 
\end{align}
Thus, we have $||\nabla f_c(y,t)||\geq m_c||y-y^*(t)||$, which implies the asymptotic convergence of $||y(t)-y^*(t)||$ according to the fact that $\lim_{t\rightarrow\infty}\nabla f_c(y,t)=0$.
\end{proof}  

\begin{remark}\label{nabla}
{Different from the design in \eqref{eq.primal} based on the fully known cost functions, Theorem \ref{centralized.oneorder} implies that, even with unknown (or inexact) parameters in the cost functions, exact optimization can still be achieved by designing an appropriate adaptive optimization algorithm.}
\end{remark}

\section{Design From a Distributed Perspective}
Inspired by the application of distributed computing in optimization problems, it is desired to design a distributed algorithm for a multi-agent system to allow them to collaboratively search for the time-varying optimal trajectory of optimization problem \eqref{optimizationproblem}.

\subsection{Design for cost Functions With Identical Hessians}
\label{Distributed Method for Single-Integrator Dynamics With Time-Varying and Identical Hessians}

The distributed algorithm includes two parts, an average estimator ({\bf C1}) and an adaptive distributed optimizer ({\bf C2}). 

{\bf C1: Average estimator.} For any $t\geq 0$, the estimator for agent $i\in\mathcal{N}$ is designed as
\begin{align}
    \dot{z}_i=&-\sum_{j\in \mathcal{N}_i}\left(\sig(\xi_i-\xi_j)^{\sigma_1}+\alpha_{ij}(t)\sgn(\xi_i-\xi_j)\right),\label{eq.z}\\
    \xi_i=&z_i+\nabla f_i(x_i,t),\label{eq.xi}\\
    \alpha_{ij}(t)=&\frac{N-1}{2}\left(\bar{\chi}_i+\bar{\chi}_j\right)+\epsilon_2,~~~~~~\epsilon_2>0,\label{eq.alphaij}
\end{align}
where $\sigma_1>1$, $\bar{\chi}_i=\bar{H}_2(||u_i(t)||_{\infty}+||x_i(t)||_{\infty})+\bar{R}$, $\forall i\in\mathcal{N}$, and $z_i\in\mathbb{R}^m$ and $\xi_i\in\mathbb{R}^m$ are estimator's internal state and output, respectively. The initial condition is given by $\sum_{i\in\mathcal{N}}z_i(0)=0$. 

{\bf C2: Distributed optimizer.} For any $t\geq{T}$, the optimizer for agent $i\in\mathcal{N}$ is designed as
\begin{align}
    u_i=&\underbrace{-\sum_{j\in\mathcal{N}_i}\beta_{ij}S(x_i-x_j,{\beta_{ij}})}_{\text{consensus coordinator}}+\underbrace{\phi_i}_{\text {optimization coordinator}},\label{eq.ui}\\
    \beta_{ij}=&\frac{N-1}{2}(||\phi_i||_\infty+||\phi_j||_\infty)+\epsilon_3,~~~~~\epsilon_3>0,\label{eq.beta}\\
    \phi_i=&-{k_1}\nabla f_i(x_i,t)-h_i^{-1}\hat{\theta}_ig_i,\\
    \dot{\hat{\theta}}_i=&N \Gamma_{\theta,i}(h_i^{-1})^{\top}\xi_ig_i^{\top},\label{eq.theta}
\end{align}
where $T>0$ is a time instant to be determined, $k_1$ is a positive constant to be determined later, $\hat{\theta}_i\in \mathbb{R}^{m\times p_i}$ is an adaptive gain matrix, and $\Gamma_{\theta,i}\in\mathbb{R}^{m\times m}$ is a positive definite constant matrix. The function $S(\cdot,\cdot)$ is defined in {\bf Notations}.

\begin{figure}[!b]
 \centerline{\includegraphics[width=0.95\columnwidth]
    {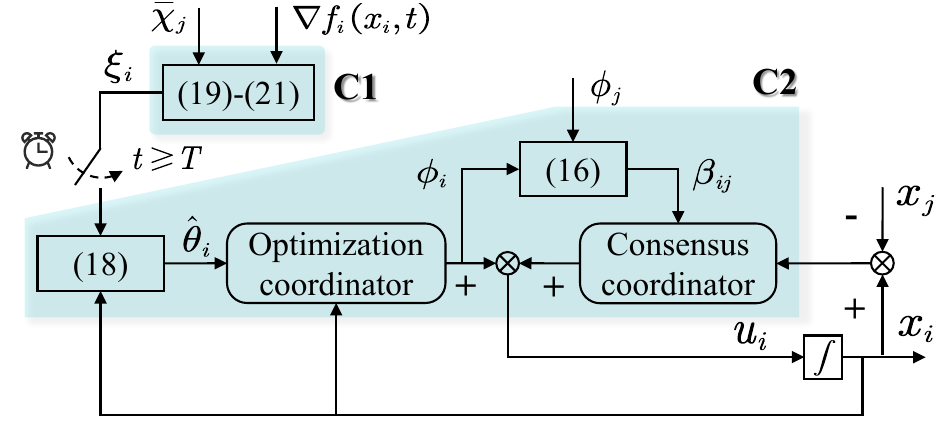}}
    \caption{The block diagram of the closed-loop system.}
    \label{fig.structure}
\end{figure}

We refer to $\alpha_{ij}$ and $\beta_{ij}$ as state-based gains as they are essentially related to the system state. {Note that the time-varying continuous approximation method\cite{Sliding98} is applied to discontinuous function in \eqref{eq.ui} and the discontinuous function only exists in \eqref{eq.z}. Then the chattering phenomenon of $x_i$ is greatly reduced. Here we introduce a nonlinear term $\sum_{j\in \mathcal{N}_i}\sig(\xi_i-\xi_j)^{\sigma_1}$ in the dynmicas of $z_i$, ensuring that the estimator converges within a given time $T$ regardless of any initial state $x_i(0)$, which is different from \cite{Ren17}. Moreover, since the continuously differentiable Lyapunov function are employed in the following analysis, we do not use the nonsmooth analysis to avoid symbol redundancy\cite{Sun23}. The block diagram of the closed-loop system is shown in Fig. \ref{fig.structure}.}
We refer to the execution logic of the proposed design as the two-stage implementation of the algorithm, because {\bf C2} needs to wait for a period of time $T$ (given by the following lemma) before starting. During this waiting time, the output $\xi_i$ of the estimator can be forced to track the dynamically changing $\sum_{j\in\mathcal{N}}\nabla f_j(x_j(t),t)/N$.
\begin{remark}
For the average estimator, the selection of state-based gain $\alpha_{ij}$ is different from that for the fixed gain in \cite{Ren17}. The reason is that the upper bound of $||(d/dt)\nabla f_i(x_i,t)||_\infty$ involves $||x_i||_\infty$. Therefore, once the implementation of the algorithm reaches the second stage, where {\bf C2} is enabled, we have no access to determine the size of $||x_i||_\infty$ and thus, the upper bound of $||(d/dt)\nabla f_i(x_i,t)||_\infty$. To address this issue, we introduce a state-based gain $\alpha_{ij}$ to guarantee the convergence of $\xi_i$. Similarly, the introduction of gain $\beta_{ij}(t)$ removes the assumption on the existence of the upper bound of $||\phi_i-\phi_j||$ in \cite{Ren17}. {Moreover, to make the design fully distributed, additional distributed consensus algorithms can be used to estimate $N$, $\bar{H}_1$, $\bar{H}_2$, $\bar{R}$ and compute the eigenvalue of $L$ \cite{FRANCESCHELLI20131031}.}
\end{remark}
Before giving the convergence analysis of the closed-loop system, we establish the convergence of estimator \eqref{eq.z}-\eqref{eq.alphaij}.

\begin{lemma}\label{lemma.finiteconverge}
Suppose Assumption \ref{assumption.graph1} holds. For system \eqref{eq.z}-\eqref{eq.alphaij}, there exists a time instant 
\begin{align}\label{timeT}
T=\frac{1}{\epsilon_2}+\frac{2}{\rho(\sigma_1-1)}>0,
\end{align}
where $\rho=\sqrt{(mN^2)^{1-\sigma_1}(2\lambda_2)^{\sigma_1+1}}$, such that $\xi_i=\chi_s(t):=\sum_{j\in\mathcal{N}}\nabla f_j(x_j,t)/N$, $\forall i\in\mathcal{N}$, $\forall t\geq T$.
\end{lemma}

\begin{proof}
    See Appendix \ref{appendix.lemma}.
\end{proof}
Then, we propose one of the main conclusions in this paper.

\begin{theorem}\label{distributed.oneorder}
Suppose Assumptions \ref{assumption.graph1}-\ref{assumption.boundedness} hold. If $H_i(t)=H_j(t)$, $\forall i,j\in\mathcal{N}$, $\forall t\geq0$, for the system \eqref{eq.multiagent} with \eqref{eq.z}-\eqref{eq.theta}, by selecting {$k_1>\sqrt{m}\bar{H}_2/(2\bar{H}_1^2)$},
the optimization problem (\ref{optimizationproblem}) is solved, i.e., $\lim_{t\rightarrow \infty} x_i(t)-x^*(t)=0$, $\forall i\in\mathcal{N}$.
\end{theorem}

\begin{proof}
  See Appendix \ref{appendix.proof1}.
\end{proof}

\begin{remark}\label{remarkbb}
Although Theorem \ref{distributed.oneorder} requires Hessians to be identical, it holds for many important cost functions, such as the generation cost function of distributed generators, $f_i(x_i,t)=a(t)x_i^2+b(t)x_i+c(t)$ with $a>0$ and $b,c\in\mathbb{R}$, and the function for optimizing electrical power supply in a power grid to meet a specific time-varying load demand, $f_i(x_i,t)=\left(x_i+k_1r_1(t)\right)^2$ with $k_1\in\mathbb{R}$. The same assumption is also adopted by \cite{Ren17} and \cite{Huang20}. 
\end{remark}

\subsection{Design for cost Functions With Nonidentical Hessians}
\label{TN}
The design in Section \ref{Distributed Method for Single-Integrator Dynamics With Time-Varying and Identical Hessians} is for a class of cost functions with identical Hessians. In fact, the average estimator can adequently estimate parameter information about the global cost functions in advance, allowing the assumption regarding identical Hessians to be relaxed. Using the two-stage implementation concept, we propose the following design.
 
{\bf D1: Average estimator.} For any $t\geq0$, the estimator for agent $i\in\mathcal{N}$ is designed as
\begin{align}
    \dot{z}_i^n=&-\sum_{j\in \mathcal{N}_i}\left(\sig(\xi_i^n-\xi_j^n)^{\sigma_2}+\alpha_{ij}^n(t)\sgn(\xi_i^n-\xi_j^n)\right),\nonumber\\
    \xi_i^n=&z_i^n+\nabla f_i(x_i,t),\label{eq.zn}\\
    \dot{z}_i^g=&-\sum_{j\in \mathcal{N}_i}\left(\sig(\xi_i^g-\xi_j^g)^{\sigma_2}+\alpha_{ij}^g(t)\sgn(\xi_i^g-\xi_j^g)\right),\nonumber\\
    \xi_i^g=&z_i^g+g_i(x_i,t),\label{eq.zg}\\
    \dot{z}_i^h=&-\sum_{j\in \mathcal{N}_i}\left(\sig(\xi_i^h-\xi_j^h)^{\sigma_2}+\alpha^h\sgn(\xi_i^h-\xi_j^h)\right),\nonumber\\
    \xi_i^h=&z_i^h+h_i(t),\label{eq.zh}
\end{align}
where 
\begin{align}
\sigma_2>&1,~~~~~~~~~~\alpha^h>(N-1)\bar{H}_2,\\
\alpha_{ij}^n(t)>&(N-1)/2\left(\bar{\chi}_i+\bar{\chi}_j\right),\label{eq.alphan}\\
\alpha_{ij}^g(t)>&(N-1)/2\left(||g_i(x_i,t)||_{\infty}+||g_j(x_j,t)||_{\infty}\right).\label{eq.alphag}
\end{align}
The initial condition for estimator is given by
\begin{align}\label{eq.conditions}
    \sum_{j\in\mathcal{N}}z_j^n(0)=\sum_{j\in\mathcal{N}}z_j^g(0)=\sum_{j\in\mathcal{N}}z_j^h(0)=0. 
\end{align}

{\bf D2: Distributed optimizer.} For any $t\geq{T}_1$, the optimizer for agent $i\in\mathcal{N}$ is designed as
\begin{align}
    u_i=&-\sum_{j\in\mathcal{N}_i}\sig(x_i-x_j)^{\sigma_3}+w_i,\label{eq.uui}\\
    w_i=&-{k_2}\xi_i^{n}-(\xi_i^{h})^{-1}\hat{\theta}_i\xi_i^g,\label{eq.wwi}\\
\dot{\hat{\theta}}_i=&N\bar\Gamma((\xi_i^{h})^{-1})^{\top}\xi_i^n(\xi_i^g)^{\top},\label{eq.thetaa}
\end{align}
where $0<\sigma_3<1$, $T_1$ is a time instant, $k_2$ is a positive constant to be determined later, and $\bar\Gamma\in\mathbb{R}^{m\times m}$ is a positive definite constant matrix. The initial condition is $\hat{\theta}_i(T_1)=\hat{\theta}_j(T_1)$, $\forall i,j\in\mathcal{N}$.  

\begin{assumption}\label{assumption.globalconvexyity}
{The function $\sum_{i\in\mathcal{N}}f_i(r,t)$ is uniformly $\bar{H}_3$-strongly convex with respect to $r$, $\forall t\geq 0$.}
\end{assumption}
\begin{theorem}\label{distributed.full}
Suppose Assumptions \ref{assumption.graph1}, \ref{assumption.boundedness}-\ref{assumption.globalconvexyity} hold. For the system \eqref{eq.multiagent} with \eqref{eq.zn}-\eqref{eq.thetaa}, by selecting {$k_2>\sqrt{m}\bar{H}_2/(2\bar{H}_3^2)$}, the optimization problem (\ref{optimizationproblem}) is solved, i.e., $\lim_{t\rightarrow \infty} x_i(t)-x^*(t)=0$, $\forall i\in\mathcal{N}$.
\end{theorem}

\begin{proof}
See Appendix \ref{appendix.proof2}.
\end{proof}
The proposed design in this section does not require the Hessians of all local cost functions to be identical. However, as a cost, it places higher demands on communication resources compared to the algorithm in the previous section, as there are three different estimators in {\bf D1} that need to exchange variables $\xi_i^n$, $\xi_i^g$ and $\xi_i^h$ with neighboring agents.  
\begin{remark}
    {For Theorem \ref{distributed.oneorder}, suppose that the initial condition for estimator {\bf C1} satisfies $\sum_{i\in\mathcal{N}}z_i(0)=\mu\neq0$. It can be concluded that $\xi_i(t)=\sum_{j\in\mathcal{N}}\nabla f_j(x_j(t),t)/N+\mu/N$, $\forall t\geq T$. Then, the convergence of $\sum_{i=1}^{N}\nabla f_i(x_i,t)$ cannot be established, but the convergence analysis of consensus error $e$ in the proof still holds. This means the optimality of $x_i$ is not robust to initial error $\mu$, but consensus of $x_i$ is always guaranteed. This conclusion also holds for Theorem \ref{distributed.full} if the initial condition \eqref{eq.conditions} is violated. To compensate for this initial error, one may use $\nabla f_i(x_i,t)-\mu/N$ to replace $\nabla f_i(x_i,t)$ in {\bf C1}, then the proof still holds.}
    \end{remark}
\section{Simulation}
\label{section.simulation}

In this section, we apply the aforementioned designs to the source seeking problem. {Note that the following controlled agents are all modeled as single-integrator dynamics.}

\subsection{Case 1}

\begin{figure}[!h]
    \centerline{\includegraphics[width=0.7\columnwidth]
    {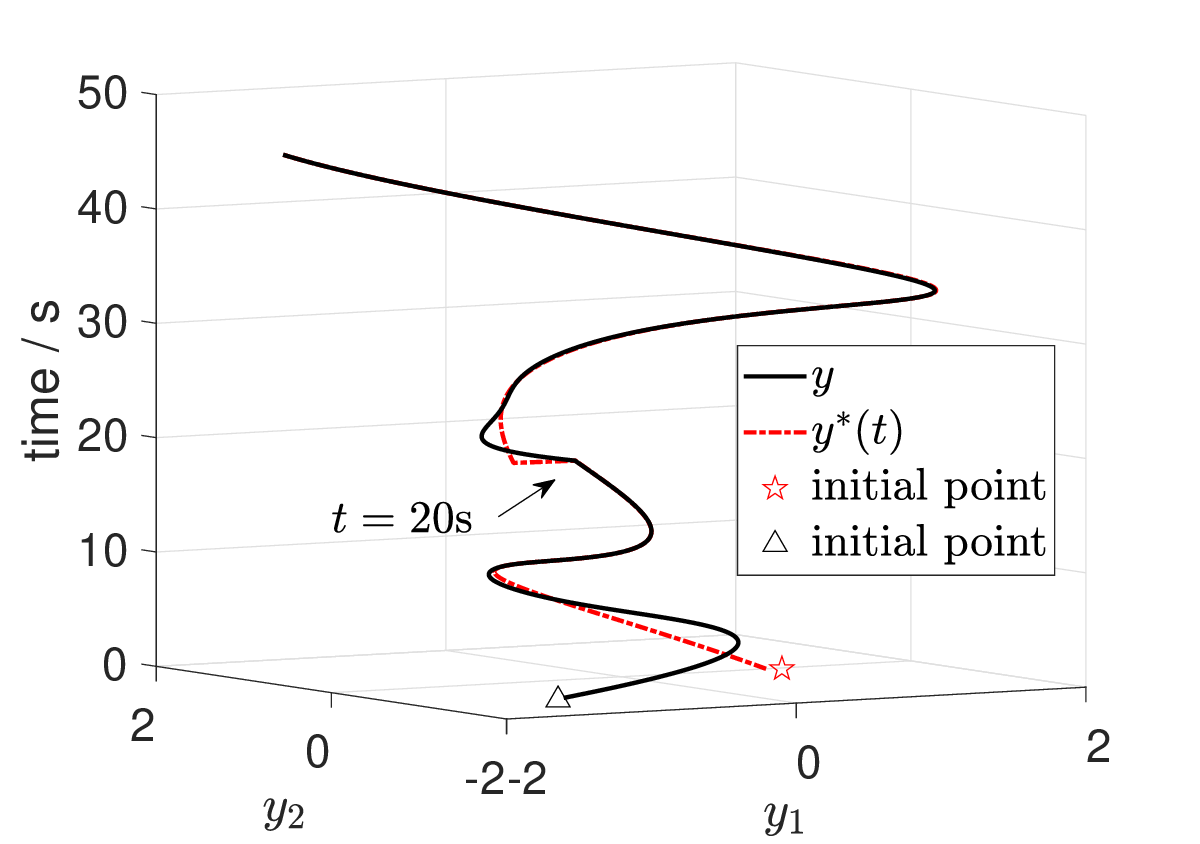}}
    \caption{The trajectories of $y^*(t)$ and $y$.}
    \label{fig.oneagentposition}
\end{figure}

\begin{figure}[!h]
    \centerline{\includegraphics[width=0.8\columnwidth]
    {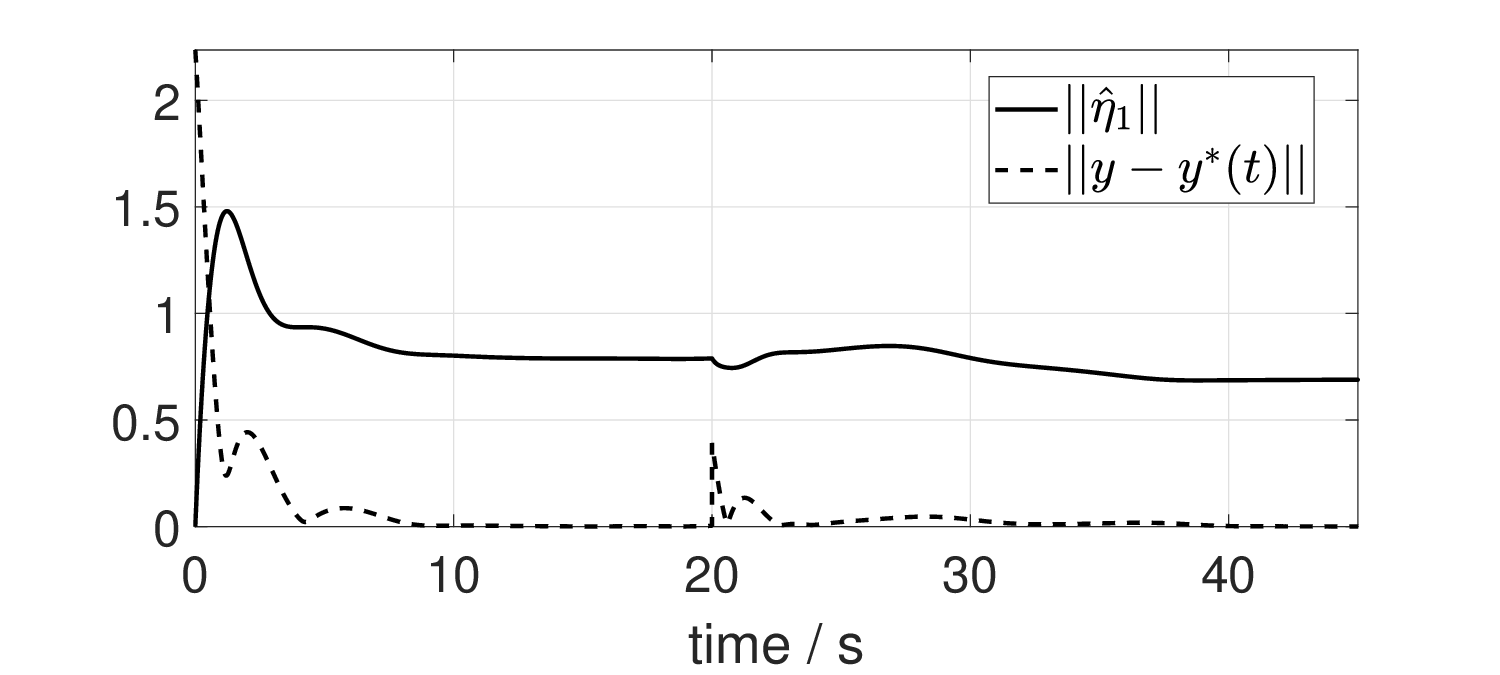}}
    \caption{The trajectories of $||\hat{\eta}_1||$ and $||y-y^*(t)||$.}
    \label{fig.normetaerror}
\end{figure}

Consider the example given by Section \ref{problem}. Let $m=2$, and the parameters of model \eqref{eq.modelstrength} are given by 
\begin{align}
  a(t)=u_1\left(1+\frac{1}{1+t}\right),~~~~r(t)=\left[ \begin{array}{c}
    u_2\cos(0.2t)\\
    u_3\sin(0.3t)\\
    \end{array} \right]. 
\end{align}
where parameters $u_1>0$ and $u_2,u_3\in\mathbb{R}$ are unknown, but the upper bounds of them are known for robots. {Then, the cost function $P^{-1}(y,t)=a^{-1}(t)||y-r(t)||^2$ has a quadratic form as shown by \eqref{eq.fi} with} 
\begin{align}
    H(t)=\frac{2(1+t)}{\mu_1(2+t)},~~~R(t)=\frac{2(1+t)}{\mu_1(2+t)}\left[ \begin{array}{c}
        u_2\cos(0.2t)\\
        u_3\sin(0.3t)\\
        \end{array} \right].\nonumber
\end{align}
{Therefore, it can be checked that the time-varying $H(t)$ and $R(t)$ here satisfy Assumptions \ref{assumption.localcostfunction}-\ref{assumption.boundedness} since that $u_1$, $u_2$ and $u_3$ are all bounded.} To solve the source seeking problem with one robot, we apply the algorithm \eqref{eq.f.controla}-\eqref{eq.f.controlb} proposed in Section \ref{cen} with $\gamma_1=0.8I_2$ and the initial sates $y(0)=[-1,-1]^{\top}$, $\hat{\eta}_1=[0]_{2\times6}$. Inspired by the fact that the parameters of the signal may change at any time\cite{kim2014cooperative}, we simulate transient changes in the signal by switching $u_2$ and $u_3$ to another set of parameters at time instant $t=20s$, which in turn leads to a `jump' in the optimal trajectory $y^*(t)$. {Fig. \ref{fig.oneagentposition} shows the trajectories of $y^*(t)$ and $y$ . Fig. \ref{fig.normetaerror} shows the boundedness of the adaptive gain matrix $\hat{\eta}_1$ and the asymptotic convergence of $y-y^*(t)$.} We observe that the autonomous robot driven by the algorithm \eqref{eq.f.controla}-\eqref{eq.f.controlb} solve the optimization problem $\min_{y}P^{-1}(y,t)$ even if there exist unknown parameters in the cost function. 

\subsection{Case 2}
\label{case2}
Consider the scenario of source seeking with multiple robots \cite{Zhang2017}. There are $N$ autonomous robots to cooperatively searching a moving acoustic source with $J$ known reference anchors around the source. Note that the reference anchors enable robots to locate their own positions in a weak GPS environment. The anchor identification matrix is defined as $Q=[q_{ij}]\in\mathbb{R}^{N\times J}$, where $q_{ij}>0$ if robot $i$ can obtain the position information of anchor $j$, and $q_{ij}=0$, otherwise. It is required that each robot moves closer to the acoustic source and the available reference anchors. Thus, the local cost function of robot $i$ can be formulated as $F_i(x_i,t)=P^{-1}(x_i,t)+\sum_{j=1}^{J}q_{ij}||x_i-R_j||_2^2$, where $R_j\in\mathbb{R}^m$ represents the position of anchor $j$, and the sum of them should be minimized, i.e., $\min_{z\in\mathbb{R}^m}\sum_{i=1}^{N}F_i(z)$. The optimal solution of this problem can be regarded as the estimation of the acoustic source's trajectory.

\begin{figure}[!h]
    \centerline{\includegraphics[width=0.34\columnwidth]
    {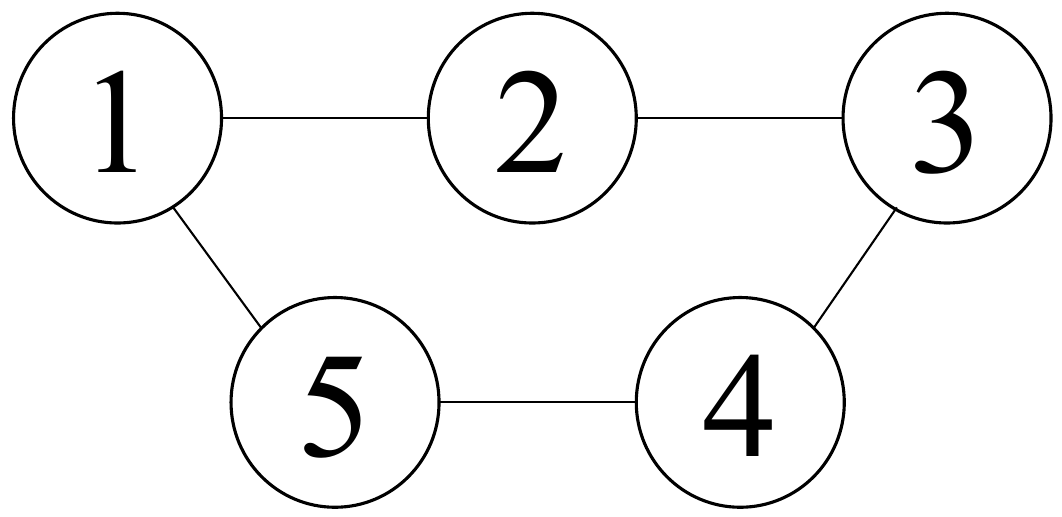}}
    \caption{Information exchange topology among robots.}
    \label{fig.topology}
\end{figure}

\begin{figure}[!h]
    \centerline{\includegraphics[width=0.7\columnwidth]
    {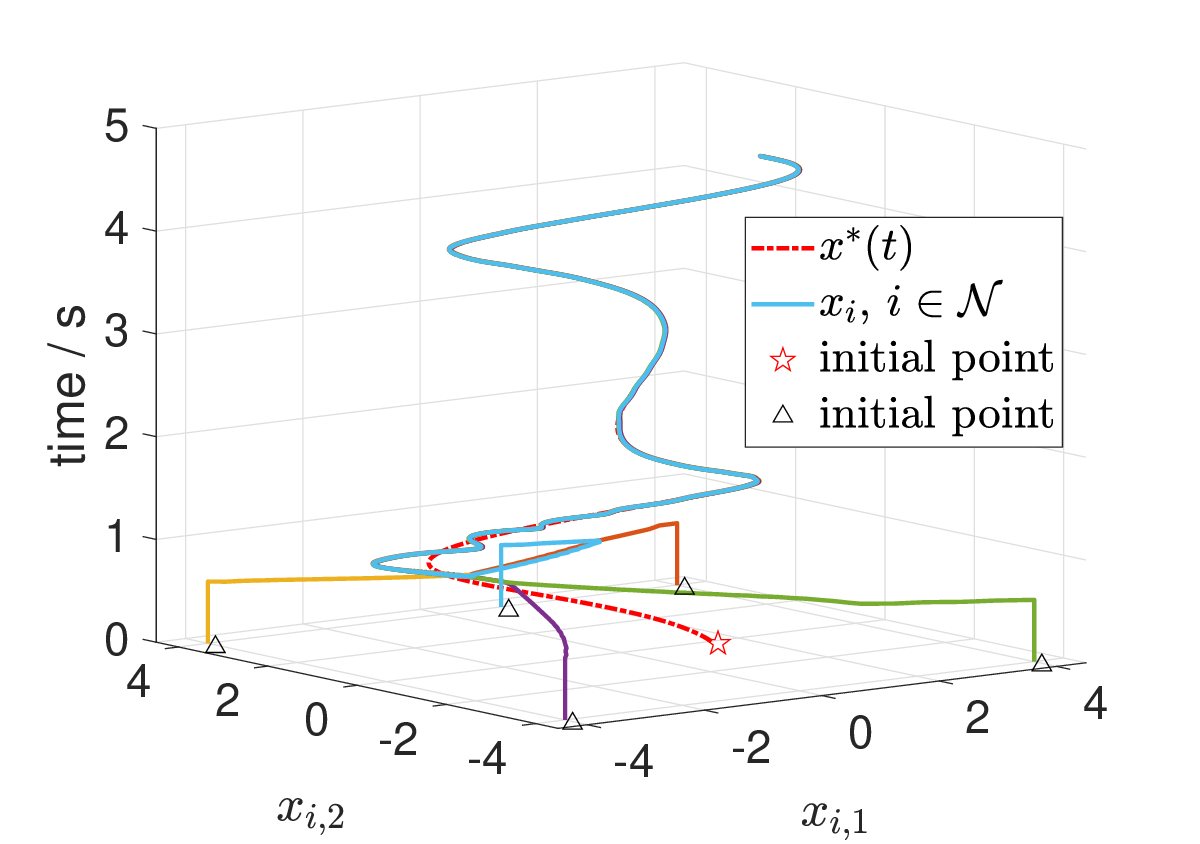}}
    \caption{The trajectories of $x^*(t)$ and $x_i$, $i\in\mathcal{N}$.}
    \label{fig.fiveagentposition}
\end{figure}

\begin{figure}[!h]
    \centerline{\includegraphics[width=0.8\columnwidth]
    {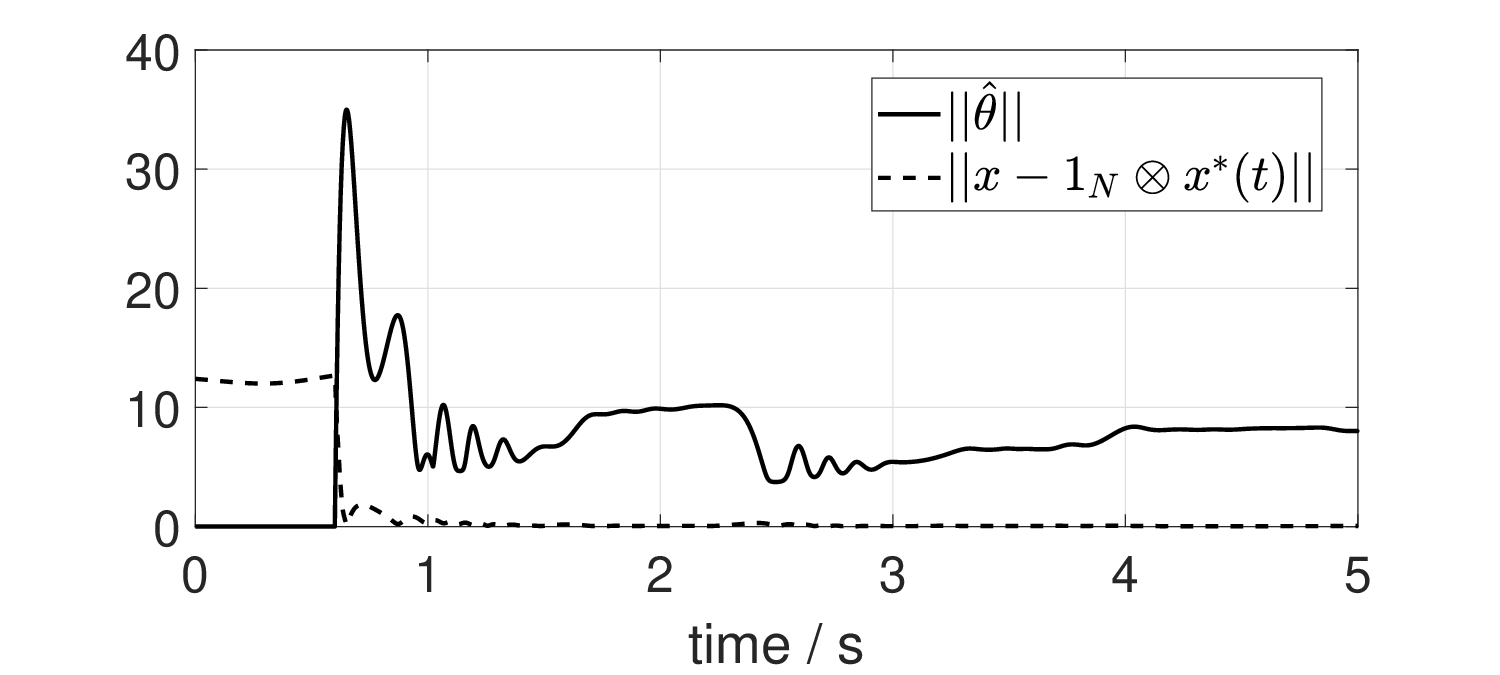}}
    \caption{The trajectories of $||\hat{\theta}||$ and $||x-1_N\otimes x^*(t)||$.}
    \label{fig.norm2}
\end{figure}

As an example, suppose that $m=2$, $N=5$, and the parameters of model \eqref{eq.modelstrength} are given by $a=u_1$, $r(t)=\left[u_2\cos(4t),u_3\sin(2.2t)\right]$, where $u_1>0$ and $u_2,u_3\in\mathbb{R}$ are not available for robots. The positions of reference anchor are given by $R_1=[-6,6]^{\top}$, $R_2=[6,6]^{\top}$, $R_3=[6,-6]^{\top}$ and $R_4=[-6,-6]^{\top}$. Let matrix $Q=0.3\times[1~1~0~0;~0~1~1~0;~0~0~1~1;$ $~1~0~0~1;~1~0~1~0]$. The information exchange topology $\mathcal{G}$ among robots is shown in Fig. \ref{fig.topology}. Then, it can be checked that Assumptions \ref{assumption.graph1}-\ref{assumption.boundedness} are satisfied and the Hessians of all cost functions are identical. We apply the design proposed in Section \ref{Distributed Method for Single-Integrator Dynamics With Time-Varying and Identical Hessians} with $\gamma_{\theta,i}=0.8I_2$, $h_i(t)=h_j(t)=I_2$, $\forall i\in\mathcal{N}$, $\epsilon_2=\epsilon_3=1$, $c=0.5$ and the initial sates $x_1(0)=[4,4]^{\top}$, $x_2(0)=[-4,4]^{\top}$, $x_3(0)=[-4,-4]^{\top}$, $x_4(0)=[4,-4]^{\top}$, $x_5(0)=[1,4]^{\top}$, $\hat{\theta}_i(0)=[0]_{2\times2}$ and $z_i(0)=[0]_{2\times1}$, $\forall i\in\mathcal{N}$. {Fig. \ref{fig.delta} shows the trajectories of $||(M\otimes I_2)\xi||$ (see Appendix \ref{appendix.lemma} for the definition of $M$), which implies that the output of estimator {\bf C1} converges to the average of the sum of the gradients within a fixed time $t=0.58$s. Fig. \ref{fig.fiveagentposition} shows the trajectories of $x^*(t)$ and $x_i$, $i\in\mathcal{N}$. Fig. \ref{fig.norm2} shows the boundedness of the adaptive gain matrix $\hat{\theta}=\col(\hat{\theta}_1,\cdots,\hat{\theta}_N)$ and the asymptotic convergence of $x_i-x^*(t)$, $i\in\mathcal{N}$. Then, the optimization problem $\min_{z\in\mathbb{R}^m}\sum_{i=1}^{N}F_i(z)$ is solved.}
\begin{figure}[!h]
    \centerline{\includegraphics[width=0.75\columnwidth]
    {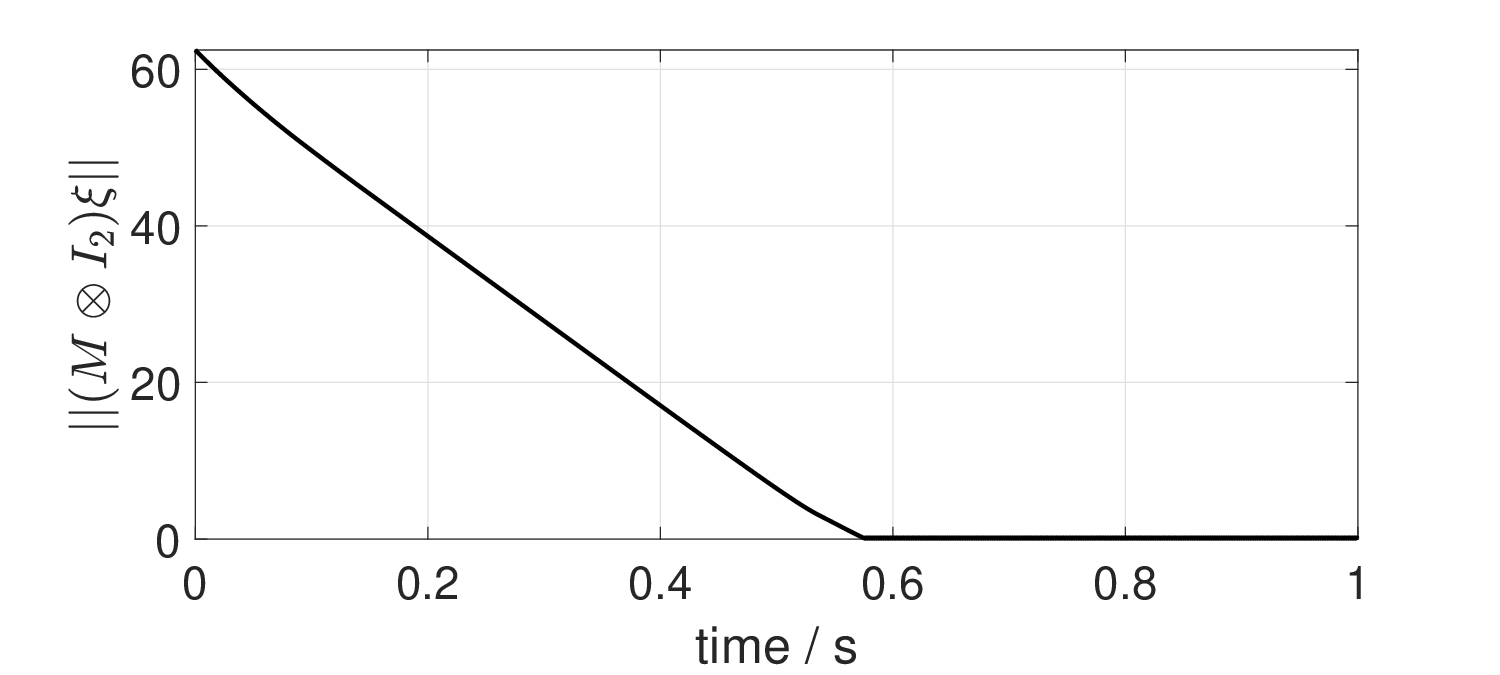}}
    \caption{The trajectory of $||(M\otimes I_2) \xi||$.}
    \label{fig.delta}
\end{figure}

\subsection{Case 3}
\begin{figure}[!h]
    \centerline{\includegraphics[width=0.75\columnwidth]
    {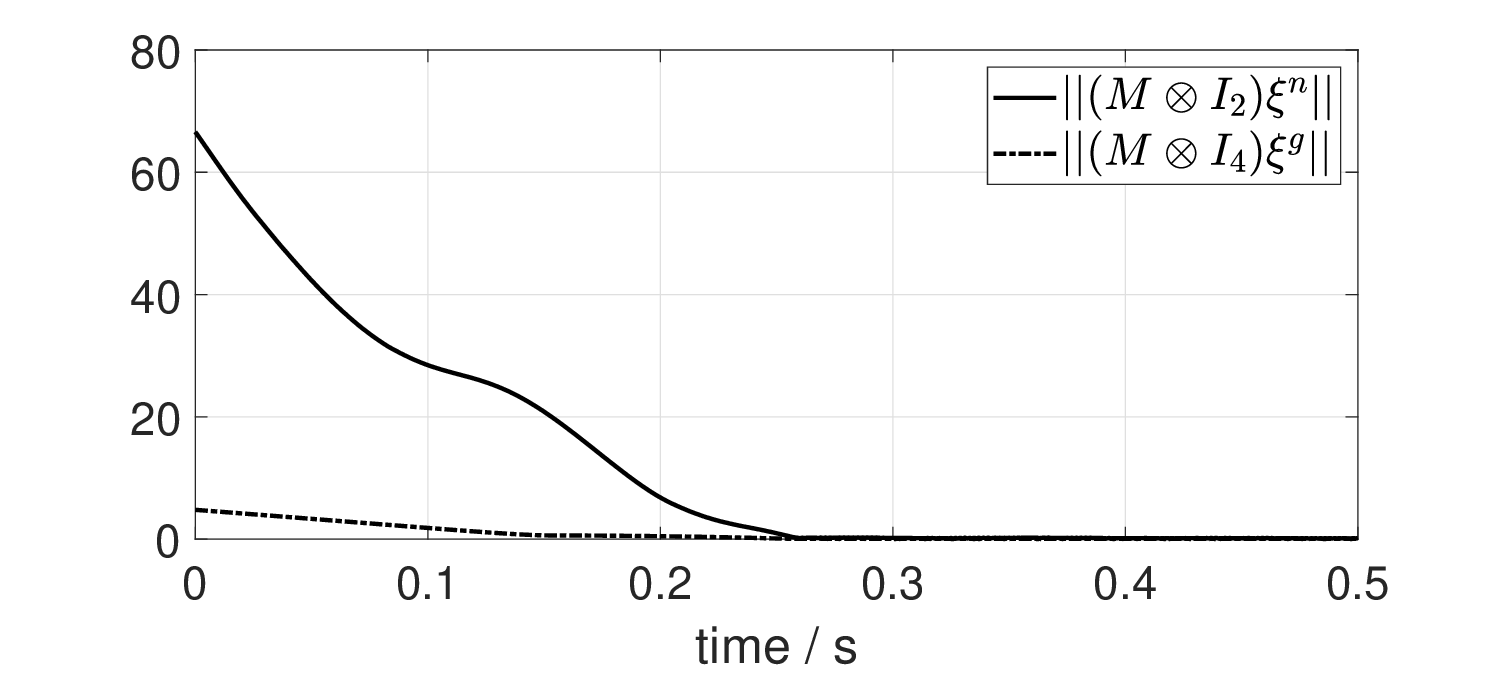}}
    \caption{The trajectories of $||(M\otimes I_2)\xi^n||$ and $||(M\otimes I_4)\xi^g||$.}
    \label{fig.mnmg}
\end{figure}

\begin{figure}[!h]
    \centerline{\includegraphics[width=0.7\columnwidth]
    {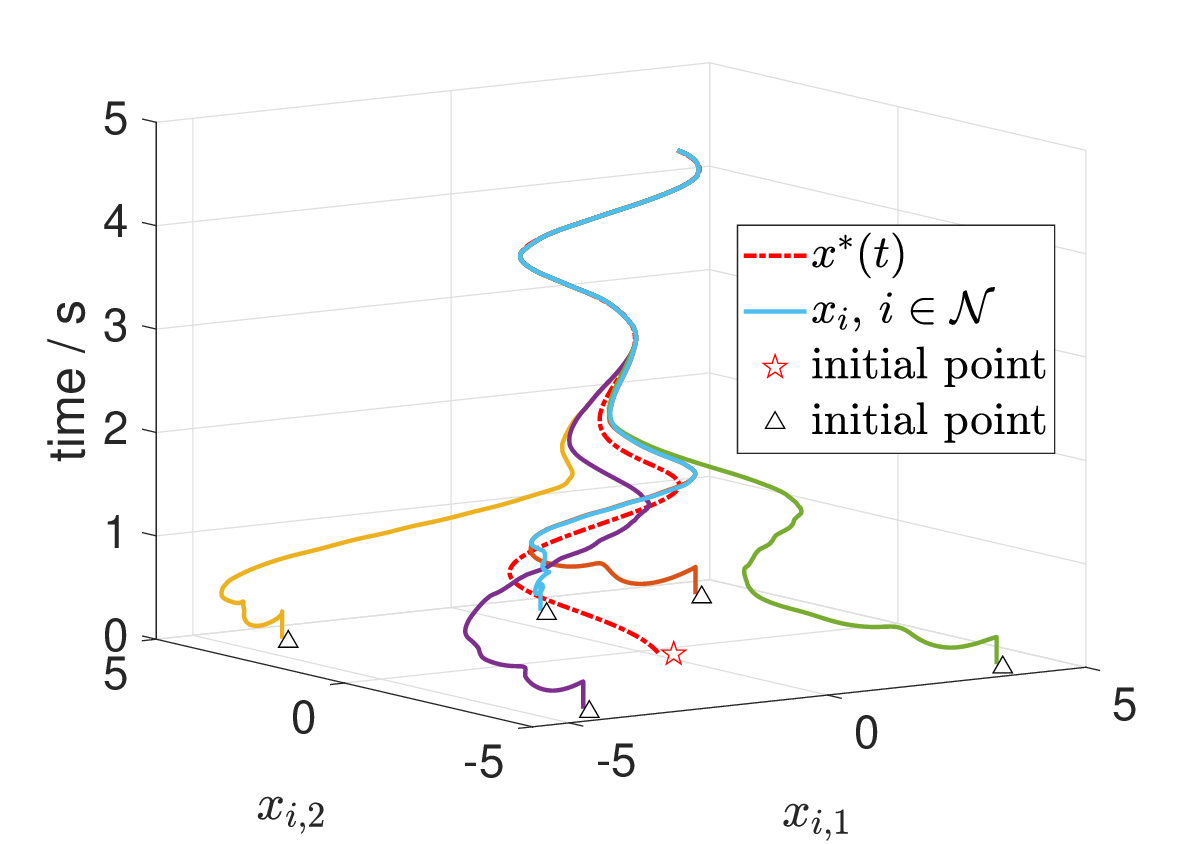}}
    \caption{The trajectories of $x^*(t)$ and $x_i$, $i\in\mathcal{N}$.}
    \label{fig.nfp}
\end{figure}

\begin{figure}[!h]
    \centerline{\includegraphics[width=0.8\columnwidth]
    {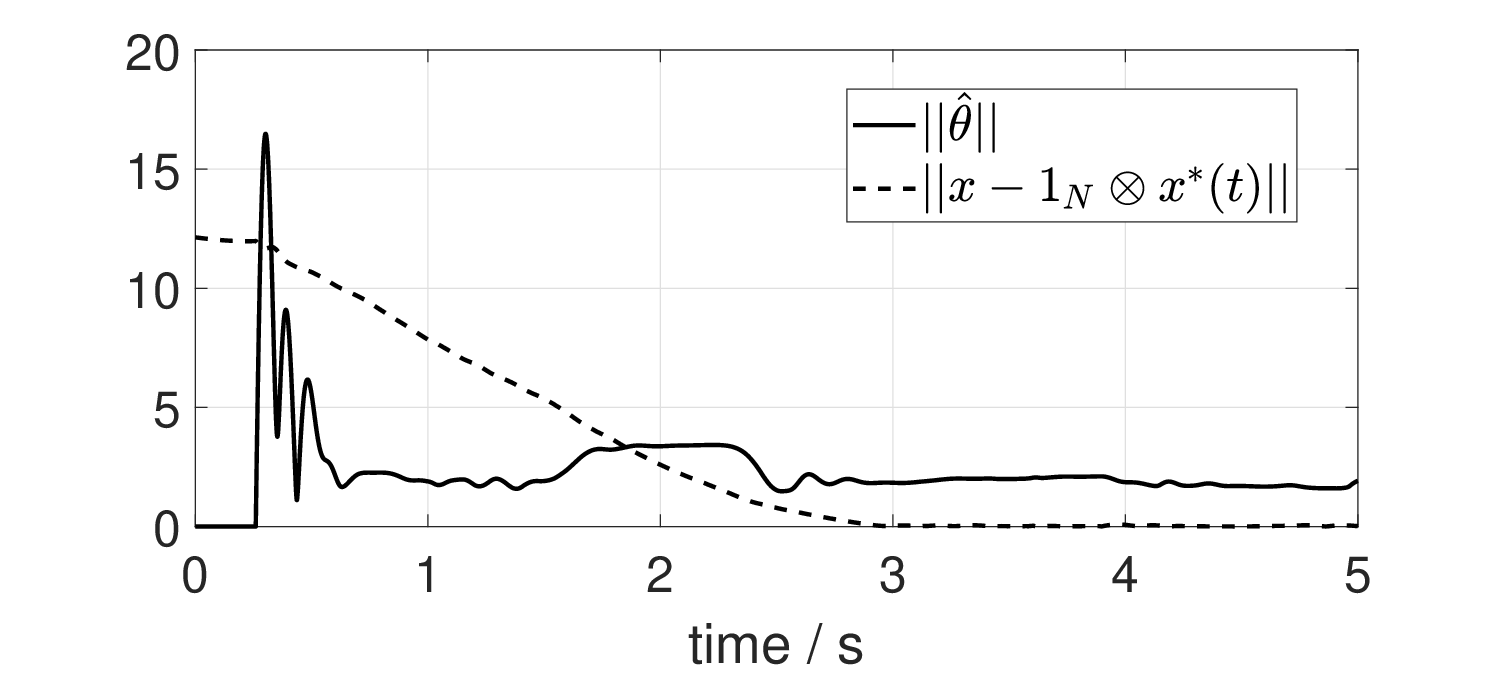}}
    \caption{The trajectories of $||\hat{\theta}||$ and $||x-1_N\otimes x^*(t)||$.}
    \label{fig.norm3}
\end{figure}

To verify the proposed design in Section \ref{TN}, consider the same scenario as Case 2 with cost functions have nonidentical Hessians. Let matrix $Q=0.1\times[3~3~0~0;~0~1~1~0;~0~0~1~3;$ $~1~0~0~1;~3~0~3~0]$. Moreover, We set the coordinates of reference anchor $\#$2 to $R_2=[\cos(2t),\sin(3t)]^{\top}$. Other parameters related to the model \eqref{eq.modelstrength}, reference anchors and information exchange topology are the same as those in Case 2. Then, it can be checked that Assumptions \ref{assumption.graph1}-\ref{assumption.boundedness} are satisfied and the Hessians of all cost functions are nonidentical. We apply the design proposed in Section \ref{TN} with $\sigma_3=0.5$ and $h_i(t)=h_j(t)=I_2$, $\forall i\in\mathcal{N}$. The initial sates are given by $\hat{\theta}_i(0)=[0]_{2\times4}$, $z_i^n(0)=[0]_{2\times1}$ and  $z_i^g(0)=[0]_{4\times1}$, $\forall i\in\mathcal{N}$. The initial position of $x_i$, $\forall i\in\mathcal{N}$ is the same as those in Case 2. {Then, Fig. \ref{fig.mnmg} illustrates the trajectories of $||(M\otimes I_2)\xi^n||$ and $||(M\otimes I_4)\xi^g||$, where $\xi^n=\col(\xi^n_1,\cdots,\xi^n_N)$ and $\xi^g=\col(\xi^g_1,\cdots,\xi^g_N)$. Fig. \ref{fig.nfp} shows the trajectories of $x^*(t)$ and $x_i$, $i\in\mathcal{N}$. Fig. \ref{fig.norm3} shows the boundedness of the adaptive gain matrix $\hat{\theta}$ and the asymptotic convergence of $x_i-x^*(t)$, $i\in\mathcal{N}$. Then, the optimization problem $\min_{z\in\mathbb{R}^m}\sum_{i=1}^{N}F_i(z)$ is solved.}


\section{Conclusion}
\label{Conclusion}
In this paper, we study the continuous-time optimization problem with uncertain time-varying quadratic cost functions. We first develop a centralized adaptive optimization algorithm using partial information of the cost function. We then extend this approach to two classes of distributed implementations for all cost functions with identical and nonidentical Hessians. The introduction of state-based gains eliminates the need for the upper bounds of complex time-varying functions. Under the mild assumption of cost functions, we prove the global asymptotic convergence of the proposed algorithms. Our work provides a theoretically feasible design for multi-robot autonomous optimization in unknown
dynamic environments. {Considering the compatibility between adaptive methods and backstepping controller design, extending the results of this paper to multi-agent systems with more general dynamics has theoretical and practical significance.}

\appendices
\section{Proof of Lemma \ref{lemma.finiteconverge}}\label{appendix.lemma}
Define a diagonal matrix $\Pi$, whose order is equal to the number of edges in the graph $\mathcal{G}$. The nodes corresponding to non-zero elements in column $i$ of matrix $D$ are denoted as $i_1$ and $i_2$, respectively. The diagonal element $\Pi_{ii}$ of $\Pi$ is defined as $\alpha_{i_1i_2}$. Then, by defining $\xi=\col(\xi_1,\cdots,\xi_N)$ and  $\chi=\col(\nabla f_1(x_1,t),\cdots,\nabla f_N(x_N,t))$, system \eqref{eq.z}-\eqref{eq.alphaij} can be rewritten as
\begin{align}
\dot{\xi}=&-(D\otimes I_m)\sig\left((D^{\top}\otimes I_m)\xi\right)^{\sigma_1}\nonumber\\
&-(D\Pi\otimes I_m)\sgn\left((D^{\top}\otimes I_m)\xi\right)+\dot{\chi}.
\end{align}
By defining consensus error $\delta_i=\xi_i-\sum_{j=1}^{N}\xi_j/N$, we have
\begin{align}
\delta=\xi-1_N\otimes \left(\frac{1}{N}\sum_{j=1}^{N}\xi_j\right)=(M\otimes I_m) \xi,
\end{align}
where $\delta=\col(\delta_1,\cdots,\delta_N)$ and $M=I_N-(1/N)1_N1_N^{\top}$. With the fact that $MD=D$, the dynamics of $\delta$ subsystem can be written as
\begin{align}\label{eq.delta}
\dot{\delta}=&-(D\otimes I_m)\sig\left((D^{\top}\otimes I_m)\delta\right)^{\sigma_1}\nonumber\\
&-(D\Pi\otimes I_m)\sgn\left((D^{\top}\otimes I_m)\delta\right)+(M\otimes I_m)\dot{\chi}.
\end{align}
Then, consider the Lyapunov function candidate 
$W=\delta^{\top}\delta$. Its derivative along \eqref{eq.delta} is given by
\begin{align}\label{eq.dotW}
\dot{W}=&-2\delta^{\top}(D\otimes I_m)\sig\left((D^{\top}\otimes I_m)\delta\right)^{\sigma_1}+2\delta^{\top}(M\otimes I_m)\dot{\chi}\nonumber\\
&-2\delta^{\top}(D\Pi\otimes I_m)\sgn\left((D^{\top}\otimes I_m)\delta\right).
\end{align}
We analyze the first term on the right side of above equation
\begin{align}\label{eq.diyige}
&2\delta^{\top}(D\otimes I_m)\sig\left((D^{\top}\otimes I_m)\delta\right)^{\sigma_1}\nonumber\\
=&-2\delta^{\top}\left( \begin{array}{c}
    \sum_{j\in\mathcal{N}_1}\sig(\delta_1-\delta_j)^{\sigma_1}\\
    \vdots\\
    \sum_{j\in\mathcal{N}_N}\sig(\delta_N-\delta_j)^{\sigma_1}\\
    \end{array} \right)\nonumber\\
=&-\sum_{i=1}^{N}\sum_{j\in\mathcal{N}_i}(\delta_i-\delta_j)^{\top}\sig(\delta_i-\delta_j)^{\sigma_1}\nonumber\\
=&-\sum_{i=1}^{N}\sum_{j\in\mathcal{N}_i}\sum_{k=1}^{m}|\delta_{i,k}-\delta_{j,k}|^{\sigma_1+1}\nonumber\\
=&-\sum_{i=1}^{N}\sum_{j\in\mathcal{N}_i}\sum_{k=1}^{m}\left(\left(\delta_{i,k}-\delta_{j,k}\right)^2\right)^{\frac{\sigma_1+1}{2}}\nonumber\\
\leq&-(mN^2)^{\frac{1-\sigma_1}{2}}\left(\sum_{i=1}^{N}\sum_{j\in\mathcal{N}_i}\sum_{k=1}^{m}\left(\delta_{i,k}-\delta_{j,k}\right)^2\right)^{\frac{\sigma_1+1}{2}},
\end{align}
where the last inequality is obtained by Lemma \ref{fangda}. Note that 
\begin{align}
\sum_{i=1}^{N}\sum_{j\in\mathcal{N}_i}\sum_{k=1}^{m}\left(\delta_{i,k}-\delta_{j,k}\right)^2=2\delta^{\top}L\delta\geq2\lambda_2\delta^{\top}\delta,    
\end{align}
where the inequality is concluded from \cite{Olfati} with the fact that $\delta^{\top}(1_N\otimes I_m)\equiv 0$. Thus, with $\rho$ given below \eqref{timeT}, we have
\begin{align}\label{eq.fixed}
2\delta^{\top}(D\otimes I_m)\sig\left((D^{\top}\otimes I_m)\delta\right)^{\sigma_1}\leq -\rho W^{\frac{\sigma_1+1}{2}}.
\end{align}
We analyze the last term on {the right side of \eqref{eq.dotW}}
\begin{align}\label{eq.dierge}
&-2\delta^{\top}(D\Pi\otimes I_m)\sgn\left((D^{\top}\otimes I_m)\delta\right)\nonumber\\
=&-\sum_{i=1}^{N}\sum_{j\in\mathcal{N}_i}\alpha_{ij}(\delta_i-\delta_j)^{\top}\sgn(\delta_i-\delta_j)\nonumber\\
=&-\sum_{i=1}^{N}\sum_{j\in\mathcal{N}_i}\alpha_{ij}||\delta_i-\delta_j||_1,
\end{align}
where the first equation is obtained by Assumption \ref{assumption.graph1}. We further analyze the second term on the right side of \eqref{eq.dotW}
\begin{align}\label{eq.disange1}
2\delta^{\top}(M\otimes I_m)\dot{\chi}=&\frac{1}{N}\sum_{i=1}^{N}\sum_{j=1}^{N}(\delta_i-\delta_j)^{\top}(\dot{\chi}_i-\dot{\chi}_j)\nonumber\\
\leq& \frac{1}{N}\sum_{i=1}^{N}\sum_{j=1}^{N}||\delta_i-\delta_j||_1||\dot{\chi}_i-\dot{\chi}_j||_{\infty}{,}
\end{align}
which is obtained by Hölder's inequality \cite{Nonlinear02khalil}. Note that
\begin{align}\label{eq.chidot}
||\dot{\chi}_i||_{\infty}\leq&||{H}_i(t)\dot{x}_i||_{\infty}+||\dot{H}_i(t)x_i||_{\infty}+||\dot{R}_i(t)||_{\infty}\leq\bar{\chi}_i. 
\end{align} 
With this in mind, by using triangular inequality $||\dot{\chi}_i-\dot{\chi}_j||_{\infty}\leq||\dot{\chi}_i||_{\infty}+||\dot{\chi}_j||_{\infty}\leq\bar{\chi}_i+\bar{\chi}_j$, we have
\begin{align}\label{eq.disange3}
2\delta^{\top}(M\otimes I_m)\dot{\chi}\leq& \frac{1}{N}\sum_{i=1}^{N}\sum_{j=1}^{N}(\bar{\chi}_i+\bar{\chi}_j)||\delta_i-\delta_j||_1\nonumber\\
\leq& \max_i\left\{\sum_{j=1,j\neq i}^{N}(\bar{\chi}_i+\bar{\chi}_j)||\delta_i-\delta_j||_1\right\}\nonumber\\
\leq& \frac{N-1}{2}\sum_{i=1}^{N}\sum_{j\in\mathcal{N}_i}(\bar{\chi}_i+\bar{\chi}_j)||\delta_i-\delta_j||_1,
\end{align}
where the last inequality is obtained by using Assumption \ref{assumption.graph1}. Substituting \eqref{eq.fixed}, \eqref{eq.dierge} and \eqref{eq.disange3} into \eqref{eq.dotW} yields
\begin{align}
\dot{W}\leq&\sum_{i=1}^{N}\sum_{j\in\mathcal{N}_i}\left(\frac{N-1}{2}(\bar{\chi}_i+\bar{\chi}_j)-\alpha_{ij}\right)||\delta_i-\delta_j||_1-\rho W^{\frac{\sigma_1+1}{2}}.
\end{align}
Note that
\begin{align}\label{eq.fangsuo}
&-\sum_{i=1}^{N}\sum_{j\in\mathcal{N}_i}||\delta_i-\delta_j||_1\nonumber\\
=&-2\delta^{\top}(D\otimes I_m)\sgn\left((D^{\top}\otimes I_m)\delta\right)\nonumber\\
=&-2||(D^{\top}\otimes I_m)\delta||_1\nonumber\\
\leq&-2\sqrt{\delta^{\top}(DD^{\top}\otimes I_m)\delta}\leq-2\sqrt{\lambda_2}||\delta||,
\end{align}
where the last inequality follows from $L=DD^{\top}$. Then, with \eqref{eq.alphaij} satisfied, we have $\dot{W}\leq-2\epsilon_2\sqrt{\lambda_2}W^{1/2}-\rho W^{(\sigma_1+1)/2}$. By using Lemma \ref{finite}, there exists a fixed time $T$ given by \eqref{timeT} such that for all $ t\geq T$, $\delta(t)=0$, i.e., $\xi_i(t)=\xi_j(t)$, $\forall i,j\in\mathcal{N}$. Moreover, by using Assumption \ref{assumption.graph1}, we have
\begin{align}
(1^{\top}_N\otimes I_m)\dot{z}=&-(1^{\top}_ND\otimes I_m)\sig\left((D^{\top}\otimes I_m)\xi\right)^{\sigma_1}\nonumber\\
&-(1^{\top}_ND\Pi\otimes I_m)\sgn\left((D^{\top}\otimes I_m)\xi\right)=0.
\end{align}
By recalling the initial condition $\sum_{i\in\mathcal{N}}z_i(0)=0$, we have $\sum_{i\in\mathcal{N}}z_i(t)=0$. Combining \eqref{eq.xi} yields
\begin{align}\label{eq.hengdeng}
\sum_{i\in\mathcal{N}}\xi_i(t)=\sum_{i\in\mathcal{N}}\nabla f_i(x_i,t),~~~\forall t\geq 0.
\end{align}
With the fact that $\xi_i(t)=\xi_j(t)$, $\forall i,j\in\mathcal{N}$, $\forall t\geq T$, we thus have $
\xi_i=\chi_s(t)$, $\forall t\geq T$.

\section{Proof of Theorem \ref{distributed.oneorder}}\label{appendix.proof1}
{\bf Analysis of average estimator.} For system \eqref{eq.z}-\eqref{eq.alphaij}, it follows from Lemma \ref{lemma.finiteconverge} that there exists a finite time instant $T>0$  such that $\xi_i(t)=\chi_s(t)$, $\forall i\in\mathcal{N}$, $\forall t\geq T$. Recalling \eqref{eq.theta}, it implies
\begin{align}\label{eq.temptheta}
    \dot{\hat{\theta}}_i&=\Gamma_{\theta,i}(h_i^{-1})^{\top}\chi_sg_i^{\top},~~\forall i\in\mathcal{N},~~\forall t\geq T.
\end{align}

{\bf Analysis of distributed optimizer.} Based on \eqref{eq.temptheta}, for any $t\geq T$, we make the following analysis.

{\em (1) Convergence of $\sum_{i=1}^{N}\nabla f_i(x_i,t)$.} Define 
\begin{align}\label{eq.dis.single.tildethetC1}
    \tilde{\theta}_i=\Omega_i^{-1}A_i-\hat{\theta}_i,~~~~~~\forall i\in\mathcal{N}.
\end{align}
With $H=\sum_{j=1}^{N}H_j$, consider a Lyapunov function candidate
\begin{align}\label{eq.v1}
    V(t)=&\frac{1}{2}\chi_{s}^{\top}H^{-1}\chi_{s}+\frac{1}{2}tr\left(\sum_{j=1}^{N}\tilde{\theta}_j^{\top}\Gamma_{\theta,j}^{-1}\tilde{\theta}_j\right).
\end{align}
The positive definiteness of $H$ can be guaranteed by Assumption \ref{assumption.localcostfunction}. Taking the derivative of $V$ yields
\begin{align}\label{eq.dis.single.dotV}
    \dot{V}=&\chi_{s}^{\top}H^{-1}\left(\sum_{j=1}^{N}H_j\dot{x}_j+\sum_{j=1}^{N}\nabla_t f_j(x_j,t)\right)\nonumber\\
    &{+\frac{1}{2}\chi_{s}^{\top}\left(\frac{d}{dt}H^{-1}\right)\chi_{s}}+\tr\left(\sum_{j=1}^{N}\tilde{\theta}_j^{\top}\Gamma_{\theta,j}^{-1}\dot{\tilde{\theta}}_j\right).
\end{align}
According to the analysis in \eqref{eq.Delta1}, we have 
\begin{align}
    {\frac{1}{2}\bigg|\bigg|\chi_{s}^{\top}\left(\frac{d}{dt}H^{-1}\right)\chi_{s}\bigg|\bigg|\leq\frac{\sqrt{m}\bar{H}_2}{2N\bar{H}_1^2}||\chi_{s}||^2=:\Delta_2.}
\end{align}
With the assumption that $H_i=H_j=:H_0$, $\forall i,j\in\mathcal{N}$, we have
\begin{align}\label{}
\dot{V}\leq&\frac{1}{N}\chi_{s}^{\top}H^{-1}_0\left(H_0\left(-{k_1}\chi_{s}-\sum_{j=1}^{N}h_j^{-1}\hat{\theta}_jg_j\right)+\sum_{j=1}^{N}A_jg_j\right)\nonumber\\
&+\tr\left(\sum_{j=1}^{N}\tilde{\theta}_j^{\top}\Gamma_{\theta,j}^{-1}\dot{\tilde{\theta}}_j\right){+\Delta_2}\nonumber\\
=&-\frac{{k_1}}{N}||\chi_{s}||^2+\frac{1}{N}\chi_{s}^{\top}\sum_{j=1}^{N}h_j^{-1}\left(\Omega_j^{-1}A_j-\hat{\theta}_j\right)g_j\nonumber\\
&+\tr\left(\sum_{j=1}^{N}\tilde{\theta}_j^{\top}\Gamma_{\theta,j}^{-1}\dot{\tilde{\theta}}_j\right){+\Delta_2}.
\end{align}
Substituting \eqref{eq.dis.single.tildethetC1} into it yields
\begin{align}
    \dot{V}\leq&-\frac{{k_1}}{N}||\chi_{s}||^2+\frac{1}{N}\chi_{s}^{\top}\sum_{j=1}^{N}h_j^{-1}\tilde{\theta}_j{g}_j+\tr\left(\sum_{j=1}^{N}\tilde{\theta}_j^{\top}\Gamma_{\theta,j}^{-1}\dot{\tilde{\theta}}_j\right){+\Delta_2}\nonumber\\
    =&-\frac{{k_1}}{N}||\chi_{s}||^2+\frac{1}{N}\tr\left(\sum_{j=1}^{N}\tilde{\theta}_j^{\top}(h_j^{-1})^{\top}\chi_{s}{g}_j^{\top}\right)\nonumber\\
    &-\frac{1}{N}\tr\left(\sum_{j=1}^{N}\tilde{\theta}_j^{\top}\Gamma_{\theta,j}^{-1}\dot{\hat{\theta}}_j\right){+\Delta_2}\nonumber\\
    =&-\frac{1}{N}{\left(k_1-\frac{\sqrt{m}\bar{H}_2}{2\bar{H}_1^2}\right)}||\chi_{s}||^2,
\end{align}
where the last equation follows from \eqref{eq.temptheta}. Then, if $k_1>\sqrt{m}\bar{H}_2/(2\bar{H}_1^2)$ holds, we have $V(t)\leq V(T)$, $\forall t\geq T$, and thus $\chi_{s}$ and $\hat{\theta}_i$, $\forall i\in\mathcal{N}$ are bounded. According to the similar analysis in Theorem \ref{centralized.oneorder}, we have $\lim_{t\rightarrow\infty}\chi_{s}=0$, i.e., $\lim_{t\rightarrow\infty}\sum_{i=1}^{N}\nabla f_i(x_i,t)=0$. Then, we will give the consensus analysis to guarantee that $x_i(t)=x_j(t)$, $\forall i,j\in\mathcal{N}$ as $t\rightarrow\infty$.

{\em (2) Convergence of consensus error.}
By defining consensus error $e_i=x_i-\sum_{j=1}^{N}x_j/N$ and its compact form $e=\col(e_1,\cdots,e_N)$, we have $e=(M\otimes I_m) x$. Then, the error dynamics is written as
\begin{align}\label{eq.e}
    \dot{e}&=-\left( \begin{array}{c}
        \sum_{j\in\mathcal{N}_1}\beta_{1j}S(e_1-e_j,{\beta_{1j}})\\
        \vdots\\
        \sum_{j\in\mathcal{N}_N}\beta_{Nj}S(e_N-e_j,{\beta_{Nj}})\\
        \end{array} \right)+(M\otimes I_m)\phi,
\end{align}
where $\phi=\col(\phi_1,\cdots,\phi_N)$. Consider a Lyapunov function candidate $W=e^{\top}e$. Its derivative along \eqref{eq.e} is given by
\begin{align}\label{eq.dotv2}
\dot{W}=&-2e^{\top}\left( \begin{array}{c}
    \sum_{j\in\mathcal{N}_1}\beta_{1j}S(e_1-e_j,{\beta_{1j}})\\
    \vdots\\
    \sum_{j\in\mathcal{N}_N}\beta_{Nj}S(e_N-e_j,{\beta_{Nj}})\\
    \end{array} \right)+2e^{\top}(M\otimes I_m)\phi,\nonumber\\
=&-\sum_{i=1}^{N}\sum_{j\in\mathcal{N}_i}\beta_{ij}(e_i-e_j)^{\top}S(e_i-e_j,{\beta_{ij}})+2e^{\top}(M\otimes I_m)\phi\nonumber\\
\leq& -\sum_{i=1}^{N}\sum_{j\in\mathcal{N}_i}\beta_{ij}\left(||e_i-e_j||_1-m\frac{1}{\beta_{ij}} \exp(-ct)\right)\nonumber\\
&+\frac{(N-1)}{2}\sum_{i=1}^{N}\sum_{j\in\mathcal{N}_i}\left(||\phi_i||_\infty+||\phi_j||_\infty\right)||e_i-e_j||_1\nonumber\\
=&-\sum_{i=1}^{N}\sum_{j\in\mathcal{N}_i}\left(\beta_{ij}-\frac{N-1}{2}(||\phi_i||_\infty+||\phi_j||_\infty)\right)||e_i-e_j||_1\nonumber\\
&+\sum_{i=1}^{N}\sum_{j\in\mathcal{N}_i}m\exp(-ct),
\end{align}
{where the second equation is obtained using Assumption \ref{assumption.graph1} and the fact that $-S(y,\epsilon_1)=S(-y,\epsilon_1)$ for any $y$ and $\epsilon_1$ given by {\bf Notations}, the first half of the third inequality follows from the fact that 
\begin{align}\label{eq.ysy}
    yS(y,\epsilon_1)=|y|-\frac{|y|\exp(-ct)}{\epsilon_1|y|+\exp(-ct)}\geq|y|-\frac{1}{\epsilon_1} \exp(-ct),
\end{align}
and the second half of the third inequality follows from the analysis in \eqref{eq.disange1}-\eqref{eq.disange3}.} Then, {by \eqref{eq.beta}}, we have
\begin{align}\label{eq.dotw}
    \dot{W}\leq& -\epsilon_3\sum_{i=1}^{N}\sum_{j\in\mathcal{N}_i}||e_i-e_j||_1+mN^2\exp(-ct)\nonumber\\
    \leq&-2\epsilon_3\sqrt{\lambda_2}W^{\frac{1}{2}}+mN^2\exp(-ct),
\end{align}
which follows the similar analysis given by \eqref{eq.fangsuo}. According to \eqref{eq.dotw}, by applying \cite[Theorem 4.19]{Nonlinear02khalil} for system \eqref{eq.ui}-\eqref{eq.theta}, it is clear that the system is input-to-state stable with the term $\exp(-ct)$ as an external input. Then, since the fact that $\lim_{t\rightarrow\infty}\exp(-ct)=0$, we have $\lim_{t\rightarrow\infty}e(t)=0$, i.e., $\lim_{t\rightarrow\infty}x_i(t)-x_j(t)=0$, $\forall i,j\in\mathcal{N}$.

Now, combining the above two-step analysis, we have $\lim_{t\rightarrow\infty} x_i(t)-x^*(t)=0$, $\forall i\in\mathcal{N}$.

\section{Proof of Theorem \ref{distributed.full}}
\label{appendix.proof2}
{\bf Analysis of average estimator.} By using the similar analysis in Lemma \ref{lemma.finiteconverge} for systems \eqref{eq.zn}-\eqref{eq.zh}, we know there exists a fixed time instant $T_1>0$ such that 
\begin{align}
 \xi_i^n=&\frac{1}{N}\sum_{j\in\mathcal{N}}\nabla f_j(x_j,t),~~~\xi_i^g=\frac{1}{N}\sum_{j\in\mathcal{N}}\nabla g_j(x_j,t),\nonumber\\
 \xi_i^h=&\frac{1}{N}\sum_{j\in\mathcal{N}}h_j(t),~~~~~~~~\forall i\in\mathcal{N},~~\forall t\geq T_1.
\end{align}
By recalling {the dynamics of ${\hat{\theta}}_i$} given by \eqref{eq.thetaa}, with the initial condition $\hat{\theta}_i(T_1)=\hat{\theta}_j(T_1)$, we have $\hat{\theta}_i(t)=\hat{\theta}_j(t)$, $\forall t\geq T_1$. This further implies that $w_i(t)=w_j(t)=:\bar{w}$, $\forall i,j\in\mathcal{N}$, $\forall t\geq T_1$.

{\bf Analysis of distributed optimizer.} For any $t\geq T_1$, we have
\begin{align}\label{eq.barw}
    \dot{x}_i=&-\sum_{j\in\mathcal{N}_i}\sig(x_i-x_j)^{\sigma_3}+\bar{w}.
\end{align}
By defining tracking error $e_{x_i}=x_i-\int_{T_1}^{t}\bar{w}\,dt$, we have 
\begin{align}
    \dot{e}_{x_i}=-\sum_{j\in\mathcal{N}_i}\sig(e_{x_i}-e_{x_j})^{\sigma_3}.
\end{align}
For the above system, according to \cite{Wang10}, there exists a finite time $T_2$ such that $e_{x_i}=e_{x_j}$, $\forall i,j\in\mathcal{N}$, $\forall t\geq T_2$, which further implies $x_i=x_j$, $\forall i,j\in\mathcal{N}$, $\forall t\geq T_2$. Thus, for any $t\geq T_1+T_2$,
\begin{align}
   \dot{x}_i=&-{k_2}\xi_i^{n}-(\xi_i^{h})^{-1}\hat{\theta}_i\xi_i^g,\label{eq.xxi}\\
   \dot{\hat{\theta}}_i=&N\bar\Gamma((\xi_i^{h})^{-1})^{\top}\xi_i^n(\xi_i^g)^{\top}.\label{eq.thetaaa}
\end{align}
By Theorem \ref{centralized.oneorder}, we have $\lim_{t\rightarrow \infty} x_i(t)-x^*(t)=0$, $\forall i\in\mathcal{N}$.

\balance
\bibliographystyle{IEEEtran}
\bibliography{IEEEabrv,bib_abb}

\end{document}